\documentclass[reqno,12pt]{amsart}
\usepackage{amsmath, amssymb, amsthm,amsfonts, bm} 
\usepackage[english]{babel}
\usepackage{bbm}
\usepackage{graphicx}
\usepackage{url}
\usepackage{epstopdf}
\usepackage[a4paper,bindingoffset=0.5cm,left=2cm,right=2cm,top=2.5cm,bottom=2cm,footskip=.8cm]{geometry}
\usepackage[pdfencoding=auto]{hyperref}
\usepackage{pdflscape}
\usepackage{rotating}
\usepackage{tikz}
\usetikzlibrary{arrows, automata,positioning,calc,shapes,decorations.pathreplacing,decorations.markings,shapes.misc,petri,topaths}
\usepackage{pgfplots}
\pgfplotsset{compat=newest}
\usetikzlibrary{plotmarks}
\usepackage{grffile}
\newlength\figureheight
\newlength\figurewidth
\pgfplotsset{%
  tick label style={font=\scriptsize},
  label style={font=\footnotesize},
  legend style={font=\footnotesize},
     every axis plot/.append style={very thick}
}
 
\usepackage{rotating}
\usepackage{amsbsy,enumerate}
\usepackage{graphicx}
\usepackage{ccaption}
\usepackage{comment}
\usepackage{mathrsfs} 
\usepackage{mathtools}
\usepackage{slashbox}
\usepackage{xcolor}
\usepackage{algorithm}
\usepackage{algpseudocode}
\usepackage{subfig}
\usepackage{multicol}
\usepackage[normalem]{ulem}

\makeatletter
\newcommand{\specialcell}[1]{\ifmeasuring@#1\else\omit$\displaystyle#1$\ignorespaces\fi}
\makeatother

\newcommand{\vb}{\vspace{3.2mm}}
\renewcommand{\hat}{\widehat}

\newcommand{\dT}{\text{d}\theta}
\newcommand{\dY}{\text{d}y}

\allowdisplaybreaks

\newcommand{\pq}{p_{{\rm quit}}}
\newcommand{\uc}{u_{\rm crit}}

\theoremstyle{plain}
\newtheorem{theorem}{Theorem}

\newtheorem{lemma}[theorem]{Lemma}

\theoremstyle{definition}

\newtheorem{example}{Example}

\theoremstyle{remark}
\newtheorem{remark}[theorem]{Remark}

\begin{document}

\title[Trusting: Alone and together]{Trusting: Alone and together}

\author[Meylahn, den Boer and Mandjes]{Benedikt V.\ Meylahn, Arnoud V.\ den Boer, and Michel Mandjes}

\begin{abstract}
We study the problem of an agent continuously faced with the decision of placing or not placing trust in an institution. The agent makes use of Bayesian learning in order to estimate the institution's true trustworthiness and makes the decision to place trust based on myopic rationality. Using elements from random walk theory, we explicitly derive the probability that such an agent ceases placing trust at some point in the relationship, as well as the expected time spent placing trust conditioned on their discontinuation thereof.

\noindent
We then continue by modelling {\it two} truster agents, each in their own relationship to the institution. We consider two natural models of communication between them. In the first (``observable rewards'') agents disclose their \emph{experiences} with the institution with one another, while in the second (``observable actions'') agents merely witness the \emph{actions} of their neighbour, {\it i.e.}, placing or not placing trust. Under the same assumptions as in the single agent case, we describe the evolution of the beliefs of agents under these two different communication models. 
Both the probability of ceasing to place trust and the expected time in the system elude explicit expressions, despite there being only two agents. 
We therefore conduct a simulation study in order to compare the effect of the different kinds of communication on the trust dynamics. 

We find that a pair of agents in both communication models has a greater chance of learning the true trustworthiness of an institution than a single agent. Communication between agents promotes the formation of long term trust with a trustworthy institution as well as the timely exit from a trust relationship with an untrustworthy institution.
Contrary to what one might expect, we find that having less information (observing each other's actions instead of experiences) can sometimes be beneficial to the agents. 
\vb

\noindent
{\sc Keywords.} trust, learning, embeddedness, social influence, institutional trust 
\vb

\noindent
{\sc Affiliations.} Korteweg-de Vries Institute for Mathematics, University of Amsterdam; Science Park 904, 1098 XH Amsterdam; The Netherlands ({\it contact}: Benedikt V.\ Meylahn {\tt\scriptsize b.v.meylahn[at]uva.nl}).
\vb

\noindent
{\sc Acknowledgments.} This research was supported by the European Union’s Horizon 2020 research and innovation programme under the Marie Skłodowska-Curie grant agreement no. 945045, and by the NWO Gravitation project NETWORKS under grant no. 024.002.003. \includegraphics[height=1em]{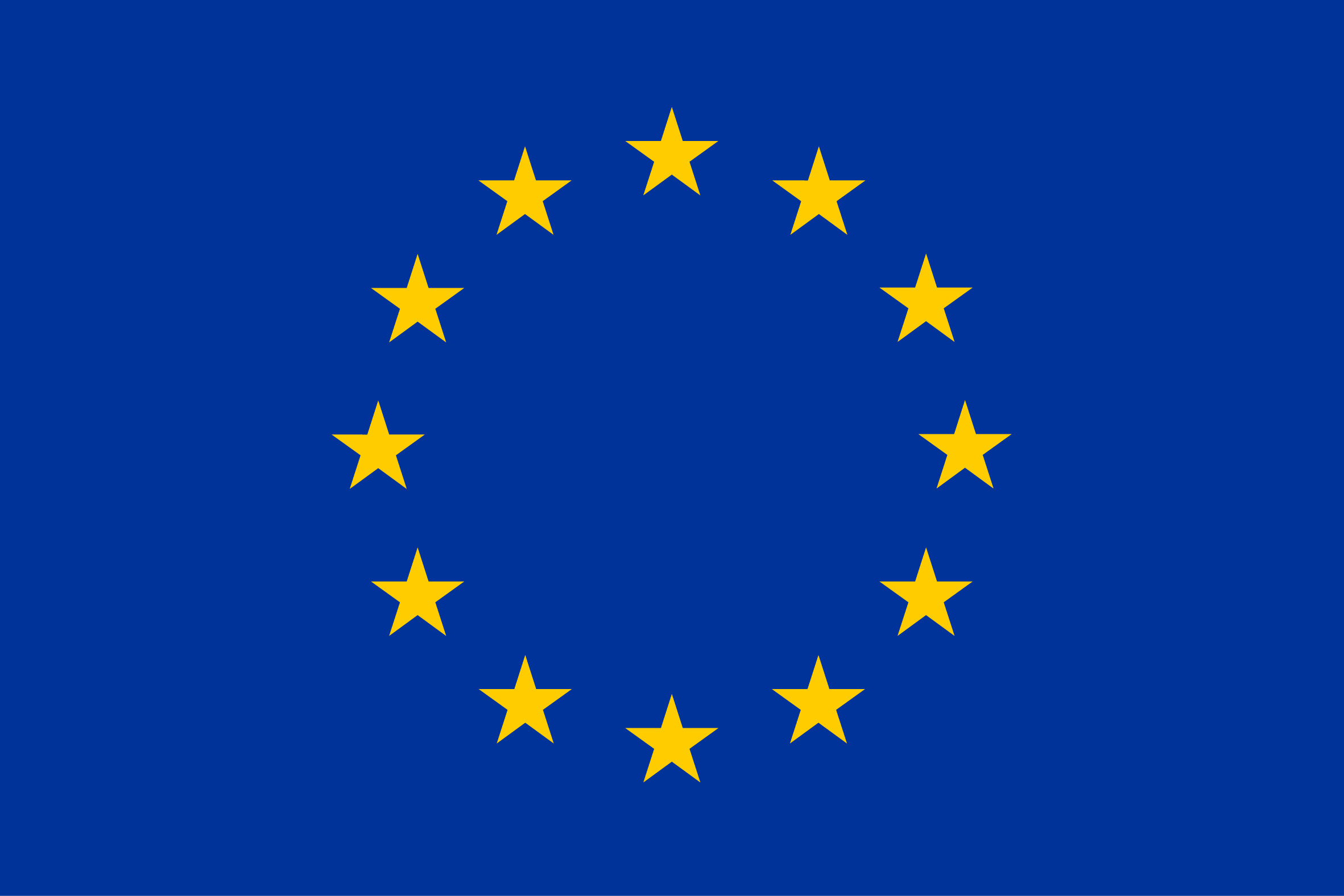}
This work was carried out on the Dutch national e-infrastructure with the support of SURF Cooperative. Version: \today.
\vb

\noindent
The authors report there are no competing interests to declare.

\end{abstract}
\maketitle
\section{Introduction}

In this paper we are interested in the process by which individuals decide to place their trust in an institution. The importance of trust and distrust in governments around the world was highlighted by the COVID-19 pandemic during which the distrust in government held by a group resulted in fewer vaccinations among individuals of that group~\cite{Bajos2022}. Similarly, distrust in science is strongly correlated with vaccine hesitancy~\cite{lazarus2022}. The prevalence of fake news during the COVID-19 infodemic~\cite{WHO2020} points toward peer-to-peer communication as a culprit for widespread distrust. Although these results were gathered in relation to the COVID-19 pandemic and vaccine hesitancy, one expects similar dynamics to apply to a whole host of different social phenomena. While the personal experience of an individual with the institution is important, it seems that the interactions with any individual's peers play a crucial role as well. We therefore study the dynamics of individual-to-institution trust in the context of peer-to-peer relationships between the individuals placing the trust.

A {\it trust situation} is a situation in which individuals are required to take a `risk' in order to observe how trustworthy the other party is~\cite[Ch.\ 1]{Buskensbook}. In some cases this risk concerns the placement of resources at the disposal of the other. Relationships to institutions are characterised by trust: the truster in such a relationship typically builds an intuition of how trustworthy the institution is. For example, we could consider the relationship between an individual and a social media company. The individual has thus placed their data (resources) in the hands of the company, and in return can post content, and has access to the posts of others. 
The level of trustworthiness the individual observes is determined by their daily experience using the social network, manifested in the (lack of) news that the social media company has been selling their information to third party advertising companies. 
As an alternative example, consider the decision of whether or not to get vaccinated for a virus. For those individuals who believe scientific consensus to have their best interest at heart, this decision may not seem like something requiring trust at all. For other individuals who are skeptical of the institution science and/or healthcare governance, it may seem like a risk to their health and possibly their personal information to get vaccinated at a government vaccine station.

This type of examples has motivated us to phrase the problem of deciding to place trust as a learning problem. The idea that agents learn about the trustworthiness of their interaction partner is also supported by sociological theory. The chapter by Buskens and Raub~\cite{Buskens2002E} provides a good overview of trust as a learning problem, as well as its own experimental evidence thereof. Furthermore, individuals are rarely isolated when faced with the decision of whether or not to place trust. Instead, human actions are embedded in a network which is likely to play a role in the decision making behaviour of its individuals. We therefore consider the effect of peer-to-peer communication on the dynamics of the trust problem, which is also supported in the empirical literature (cf. \cite{Pechmann2002, Ozdemir2020}). In particular Wang and Yu~\cite{Wang2017} find evidence that this influence acts both when experiences are communicated as well as when agents simply observe another's behaviour. We restrict our study to two agents as this showcases the essence of the underlying mechanism in the agent-to-agent communication. This allows us to construct the model with exact Bayesian learning\footnote{Work in biological perception has shown that humans often behave in a Bayesian manner. For a discussion on the implications hereof and a list of results we refer to Knill and Pouget~\cite{Knill2004}.}. Under basic assumptions of myopic\footnote{Not only is myopic decision making a common assumption in the literature (cf. \cite{Sebenius1983,Parikh1990,BalaGoyal1998,Keppo2008,Harel2021}), but there is also some experimental evidence that points to semi-myopic behaviour in humans~\cite{Yu2013,Zhang2013}.} rationality, instead of exogenously imposing model assumptions on the effect of a signal sent from one agent to the other. This approach leads to delicate intricacies already in the two agent case, stemming from agents interpreting their neighbours actions whilst knowing that these actions were influenced by the actions they themselves have taken thus far. As a result of this mutual influence, in the two agents setup the evaluation of key quantities in our model eludes simple and explicit expressions. In fact, evaluating them requires many numerical integrations, which has led us to use a cluster of a large computing facility in our numerical experiments. 

\subsection{Related literature}
Our investigation connects to two streams of literature: that of {\it social network learning} and that of {\it social network trust}. We proceed by providing an admittedly non-exhaustive, brief account of the related literature. The work relevant to our investigation spans various research disciplines, each with their own approach and set of ``reasonable assumptions.'' The fields in question contain, but are not limited to, economics, theoretical sociology, socio-physics and machine learning. Beyond modelling of these dynamics, the empirical investigations of network influence on trust even extend to the field of marketing. 

\subsubsection{Social network learning}
Social network learning concerns agents who try to optimize an unknown objective function by choosing an action, and that are learning via private signals and/or signals from other agents. Under various modelling decisions, the interest is in the speed at which a group of agents learns the best course of action, as well as whether or not under assumptions of rationality, the group may take a sub-optimal action often as a result of social influence. The modelling decisions relate to the degree of rationality in learning, the communication between the agents, and whether or not the private signals of the agents are conditional on the actions they take. Table~\ref{tab:lit} summarises the modelling decisions of the papers~\cite{BalaGoyal1998,Molavi2018,Harel2021,Huang2022}, all of them considering a population of agents in the social network learning framework. The second last column headed ``Communication'' refers to what the agents share with one another. This could be their private learning signal per round, their belief distribution, or simply the actions they take. The last column headed ``Signals'' refers to whether the reception of private signals by the agents is independent of or conditional on the actions the agents take.
\begin{table}[htb]
  \centering
  \begin{tabular}{l|lll}
    Paper & Rationality & Communication & Signals\\
    \hline
    Bala and Goyal~\cite{BalaGoyal1998} & myopic & actions and signals & conditional \\
    Molavi \textit{et al.}~\cite{Molavi2018} & imperfect-recall & belief & independent \\
    Harel \textit{et al.}~\cite{Harel2021} & myopic & actions & independent \\
    Huang \textit{et al.}~\cite{Huang2022} & non-myopic & actions & independent \\
    \hline
  \end{tabular}
  \caption{Modelling decisions in selected social network learning literature.}
  \label{tab:lit}
\end{table}

The agents in the paper of Bala and Goyal~\cite{BalaGoyal1998} are myopic, and in addition they do not infer anything about the outcomes of the neighbours of their own neighbours based on the actions taken by their immediate neighbours. The authors find that in connected networks the agents' behaviour converges asymptotically. Molavi \textit{et al.}~\cite{Molavi2018}, Harel \textit{et al.}~\cite{Harel2021} and Huang \textit{et al.}~\cite{Huang2022} all consider models in which the private signals received by the agents are independent of the actions they take. Within that context, Molavi \textit{et al.}~\cite{Molavi2018} show how imperfect-recall in the context of Bayesian learning relates to the seminal model by DeGroot~\cite{DeGroot1974}. In models of a fully connected population of agents, Harel \textit{et al.}~\cite{Harel2021} and Huang \textit{et al.}~\cite{Huang2022} relate the learning rate of the model with communication restricted to actions to a reference model where communication between agents is open to signals and beliefs.

Departing from the network setting, Correa \textit{et al.}\,\cite{Correa2020} and Fu and Le Riche~\cite{Fu2021} consider a model in which sequentially arriving agents from a population choose to buy a product or not, based on a Bayesian belief on the product's \textit{quality} that has been built on the observations of the preceding agents. Correa \textit{et al.}\,\cite{Correa2020} compute the probability of incomplete learning which occurs when the agents falsely believe the product's quality to be low. An important difference between \cite{Correa2020} and our setup is that the agents in \cite{Correa2020} know that the quality of the product comes from a set of two known values. This mean that in \cite{Correa2020} the agents are only tasked with identifying which of the two values the quality takes, whereas in our paper the relevant parameter (in the sequel refrred to as the \textit{trustworthiness}) may take any value in the unit interval. Fu and Le Riche~\cite{Fu2021} incorporate a similar learning problem into an endogenous growth market model in which the agents compare a new product of unknown quality to an old one of known quality, leading to the outcome that there are attainable equilibria in which the product's true quality remains unknown.

There also exists a body of literature focused on conditions which ensure social learning. In these papers agents typically receive only one signal and take sequential actions which are observed by some or all of the remaining agents. In the case of Banerjee and Fudenberg~\cite{Banerjee2004}, the observation structure takes the form of a representative sample while the population is a continuum. Acemoglu \textit{et al.}~\cite{Acemoglu2011} impose a network structure to determine which actions are observed by agents. For an overview of models in which agents are represented by a fully connected graph, the interested reader is referred to Birkchandani \textit{et al.}~\cite{Bikhchandani1998}.

To the best of our knowledge, the following relevant setup has not been studied in the social network learning literature: A model in which agents act myopically rational, observe their neighbours actions or signals, and have private signals which are conditional on the actions taken. This kind of model also relates to the social network trust literature, considering that trust relationships require a dependency between signals and actions.

\subsubsection{Social network trust}
Social network trust typically concerns the behaviour of pairs of truster and trustee agents playing an iterated trust game. The interest here lies in the likelihood of trust breaking down and describing the rational strategies of the truster and the trustee. Particularly relevant to our work are studies in which there is a population of trustworthy trustee agents who always honour trust, and opportunistic trustee agents who play strategically. In such models, it is a natural extension to have agents learn about the proportion of trustworthy trustee agents.

Important examples of this literature stream have been presented in Bower \textit{et al.}~\cite{Bower1996}, Buskens~\cite{Buskens2003}, and Frey \textit{et al.}~\cite{Frey2015}. Bower \textit{et al.}~\cite{Bower1996} describe equilibrium strategies for a sequence of 2-round trust games in which a new pair of truster and trustee agents is paired to play two rounds of the trust game. Additionally to learning about the true portion of trustworthy trustees, there is learning from the first round to the second round within a match-up between truster and trustee agent. Buskens~\cite{Buskens2003} extends this analysis by considering a pair of truster agents that are both in a relationship to the same trustee. The main finding is that trust placement and honouring (in equilibrium strategies) increases with the probability of sharing information between truster agents, only if both agents are sharing information with a high probability. These findings have been corroborated by experimental work by Buskens \textit{et al.}~\cite{Buskens2010}. Frey \textit{et al.}~\cite{Frey2015} extend the theoretical work further by letting the link between truster agents be bought at a cost. The authors determine which game parameters include and exclude investments in such a connection for equilibrium strategies. 

In contrast to this stream of literature, we consider a trustee agent such as an institution whose behaviour is not modelled strategically, and who interacts with more than one truster agents. Following
{\it e.g.}~\cite{Buskens2003,Frey2015,Buskens2010}, one trustee interacting with two trusters is a natural setting to consider.

\subsubsection{Relevance and perspective}
In this section we elaborate on the relevance of the two different streams of literature on the model we present. Thematically we are based in the social network trust literature, while methodologically our approach bears resemblance to those used in social network learning. We first present the ideas in both streams and subsequently how our work fits into this landscape.

The social network trust literature offers detailed descriptions of the interactions between strategic trusters and one of two types of trustees, e.g.\ ``friendly'' trustees who always honor trust and ``strategic'' trustees who try to ``fool'' the truster in order to be able to abuse trust at some point (cf.\,\cite{Bower1996, Buskens2003, Frey2015}). The proportion of ``friendly'' trustees may be known (as in~\cite{Buskens2003, Frey2015}) or unknown (as in~\cite{Bower1996}).

The trusters are learning about what type of trustee they are interacting with, and only sometimes also learning about the proportion of ``friendly'' trustees. As soon as a trustee abuses trust once in the peer-to-peer setting, they reveal that they are not to be trusted. A prevalent example in this literature is that of people buying a second hand car (cf.\,Buskens~\cite[Ch.\ 6]{Buskensbook}) from a loose acquaintance or via an online peer-to-peer service, or that of hiring an informal house sitter without a contract, both situations in which the truster and the trustee are peers in some way.

In the social network learning literature, there is an environment which creates an ordering on the actions the agents may take and emits a signal. The agents take actions and receive a signal. They use the signal to update their belief about the environment with the goal of taking the best action. It is social learning in the sense that the agents communicate with one another about their actions and/or signals. The modellers in turn are interested in the probability of learning the best action as a group (cf.\,\cite{BalaGoyal1998,Correa2020,Fu2021}) and the speed of convergence to this best action under different forms of communication (cf.\,\cite{Harel2021, Huang2022}). The social network learning results are applicable to situations in which agents are consistently updating their belief and trying to take optimal actions. One can think of prediction in markets (and setting the corresponding price to maximise profit), adoption of opinions (trying to fit in the group), and the dissemination of information (keeping up to date with the latest information).

We are motivated by the question of trust in institutions and the influence of peer-to-peer communication thereupon. Thus the social network trust literature paints the thematic backdrop for our work. We are interested in the event that trust is lost, primarily in the ``asymmetric'' case of trusters interacting with \textit{institutions}. In this setting it is less natural to incorporate the strategic behaviour of the trustee: they are not actively taking part in the interactions, but rather ``passively'' providing a service. Instead of modelling trustee behaviour as strategic, we let them simply draw an action from a distribution.

We investigated the social network learning literature because of the technical similarity between our model and the models found therein. Our goal was to look for work that handles our setting (possibly under different nomenclature). In the process we observe that our model also fills a gap in the social network learning literature. Furthermore, we are encouraged in our decision to compare the effects of different forms of communication on outcomes. In the social network trust literature, the agents are assumed to always communicate their experiences or their belief distribution fully. One can argue, though, that communication of experiences does not necessarily provide the right perspective: typically actions are readily observable, while the internal belief or personal experience may not always be shared. This has motivated us to compare the two.

The resulting modelling framework naturally applies to the setting we are interested in: individuals trusting institutions in the context of peer-to-peer influence.

\subsection{Contribution}
Our model aims to fill the above mentioned gaps in the literature.
In order to model the effect of peer-to-peer communication on the dynamics of a trust problem between an individual and an institution, we draw from both streams of literature. Learning signals are dependent on the actions taken, following the social network trust literature. We follow the social network learning literature by implementing myopically rational decisions and a Bayesian learning procedure. We study two of the communication models between agents seen in both streams related literature. In the first (``observable rewards'') agents disclose their experience with the institution with one another, while in the second (``observable actions'') agents merely witness the actions of their neighbour, {\it i.e.}, placing or not placing trust. In both models of communication, we describe rational usage of the information in updating beliefs. The extent of the rationality plays a role as a benchmark rather than a description of actual human behaviour. The computations involved become complicated quickly, but provide a useful indication of perfectly rational information usage in terms of belief updating. 

We impose no assumptions on the motivations of the trustee, who simply acts honourably at some probability $\theta\in (0,1)$, thus generalising the setting where this probability takes one of two values as in~\cite{Correa2020}. 
We are interested in a setup in which truster agents interact with institutions whose behaviour cannot be modelled strategically at the level of individual interactions, unlike~\cite{Bower1996,Buskens2003,Frey2015,KolbMadsen2022}. We also consider a more general version of the communication seen in~\cite{Buskens2003,Frey2015}, in which agents have access to their neighbours' actions \textit{and} the outcomes thereof.

We describe the information usage and myopic decision making without depending on signals independent of the actions taken as seen in~\cite{Molavi2018,Harel2021,Huang2022}. The asymmetry of the actions, natural to the trust problem, means that incomplete learning is possible. Thus we pay attention to the probability of convergence to the truth rather than only the rate thereof.

In our work we analyse the dynamics of the single agent case using techniques from the field of random walk theory. Such analytic techniques, however, do not extend to two agent models, so that we analyze these relying on Monte Carlo simulation. As the numerical integrations required for agents to interpret each others actions take a prohibitive amount of computation time on personal computing machines, we have to use the Lisa cluster of the computing facility SURFsara.

We observe that two agents with communication between them tend to make the ``correct'' decision sooner. In other words: typically, sample paths in which the relationship helps the agents outweigh the sample paths in which ``bad luck'' for one agent implies ``bad luck'' for both due to the communication between them. 

Our experiments reveal that
the observable rewards model is not always ``better'' than the observable actions model. Which mode is most helpful to the agents depends on whether one is interested in making the correct decision in the long run or in being sure to end a relationship with an untrustworthy institution as quickly as possible. Moreover, we identify a parameter setting in which the probability of quitting is \emph{lower} in the observable actions model than in the observable rewards model. This means that, contrary to what one might expect, having less information available might be beneficial to the agents, and there is no monotone ordering between the two models.


\subsection{Organisation of paper}
In \S\ref{BaseModel} we describe basic elements of the model along with relevant interpretations. Thereafter, in \S\ref{sec:SinglePlayer} the model is described further and analysed in the case of a single agent (with proofs being provided in Appendix~\ref{AX:proofs}). Here we pay special attention to subcases which allow for analytic results regarding the probabilities of such a trust relationship ceasing. In \S\ref{sec:2Player} we formulate the two models for two agents in the trust relationship with the same institution: observable rewards and observable actions. In \S\ref{sec:Experi} we discuss the experimental setup used to investigate the two agent model and in \S\ref{sec:res} we discuss the results of this experimentation. We conclude this paper with a discussion on the implications of the results and their relevance to the present literature in \S\ref{sec:disc}.

\section{Model for a single agent}\label{BaseModel}
In this section we present the model for a single agent, which forms the core of the two agents models discussed in later sections. We consider the situation in which an agent has repeated opportunities to place trust in a institution. The institution's behaviour is modelled by a single parameter $\vartheta \in [0,1]$, the true {\it trustworthiness}, defined as the probability at which trust is honoured (so that its complement $1-\vartheta$ is the probability that trust is abused). If trust is not placed, then the institution has no action to take. This behaviour can be interpreted as the efficacy of the institution honouring trust, implicitly assuming that this is what they are attempting in each round. Note that we acknowledge the high level of abstraction taken in regard to the institution and that interest is mainly in the agent's behaviour in such a situation. In each round $t\in\mathbb{N}$ the agent chooses an action from the action set $\mathcal{A}=\{0,1\}$ in which $A_t=1$ indicates that the agent places trust in round $t$ while $A_t=0$ indicates that the agent quits the trust relationship. We define the random variable $X_t$, $\forall t\in\mathbb{N}$, indicating whether the institution's action in round $t$ is (would be) that of honouring or abusing any trust that may or may not have been placed:
\begin{equation}
 X_t = 
  \begin{cases}
  1, \quad\text{with probability }\vartheta \\
  0, \quad\text{with probability }1-\vartheta.
  \end{cases}
\end{equation}
The random variables $\{X_t : t \in \mathbb{N}\}$ are i.i.d., in particular independent from the agent's actions. Importantly, the agent only observes $X_t$ in rounds when $A_t=1$. If trust is honoured the agent gains utility $r$ (reward), while if trust is abused the utility gain is $-c$ (cost). We assume that $c$ and $r$ are positive integers; note that if $r,c\in\mathbb{Q}$ we simply multiply both by the product of their denominators to get integers. The utility for the agent placing trust in the institution is $rX_t\mathbf{1}_{\{A_t=1\}}-c(1-X_t\mathbf{1}_{\{A_t=1\}})$, so that the expected utility is $r\vartheta-c(1-\vartheta)$. As is widely adopted in the learning literature (see, e.g., Sebenius and Geanakoplos~\cite{Sebenius1983}, Parikh and Krasicki~\cite{Parikh1990}, Bala and Goyal~\cite{BalaGoyal1998}, Keppo \textit{et al.}\,\cite{Keppo2008} and Harel \textit{et al.}~\cite{Harel2021}), we let the agent act with myopic rationality. This means that in every round $t$ they only consider the expected utility of the immediate action, and they do not take into account the possible utility of actions taken in rounds $t+1,t+2,\ldots$ or the utility of information gained by taking action $A_t=1$. The agent places trust if they \textit{believe} the utility to have a nonnegative expected value. Furthermore the agent starts with a Beta distributed prior belief $P_0$ with parameters $\alpha$, $\beta\in\mathbb{N}$, such that they initially believe the probability density of $\vartheta$ is given by
\begin{equation}\label{eq:belief_1pl}
  P_0(\theta)=B(\theta;\alpha,\beta):= \frac{\theta^{\alpha-1}(1-\theta)^{\beta-1}}{\int_0^1 y^{\alpha-1}(1-y)^{\beta-1} \dY}, \quad \theta \in [0,1].
\end{equation}
The initial estimate of the expectation of $\theta$ is $\hat{\vartheta}_0=\mathbb{E}[B(\alpha,\beta)]=\alpha/(\alpha+\beta)$. As more information becomes available (trust is placed and subsequently honoured or abused during rounds $t>0$) the agent updates this belief distribution in a Bayesian fashion. Let 
\begin{equation}
  \hat{S}_t=\sum_{s=1}^{t}X_s \mathbf{1}_{\{A_s=1\}},
\end{equation} 
be the number of times that the agent observes that trust was honoured until time $t$. Similarly let
\begin{equation}
\hat{F}_t = \sum_{s=1}^t(1-X_s)\mathbf{1}_{\{A_s=1\}},
\end{equation}
be the number of times that the agent observes that trust was abused until time $t$. The belief distribution held by the agent at the end of round $t$ is then found by applying Bayes' rule, so as to obtain
\begin{equation} 
  P_{t}(\theta) = \frac{\theta^{\hat{S}_t}(1-\theta)^{\hat{F}_t}P_0(\theta)}{\int_0^1 y^{\hat{S}_t}(1-y)^{\hat{F}_t}P_0(y)\dY}, \quad\theta \in [0,1].
\end{equation}

 Denoting the estimated trustworthiness at time $t$ by 
 \begin{equation}
   \hat{\vartheta}_t=\mathbb{E}_{\theta\sim P_t(\theta)}[\theta] = \frac{\alpha+\hat{S}_t}{\alpha+\beta+\hat{S}_t+\hat{F}_t}, \quad\forall t\in \mathbb{N},
 \end{equation}
 then for $t=1,2,\ldots$ we have:
\begin{equation}\label{eq:Action}
  A_t=\begin{cases}
  1, \quad\text{if }r\hat{\vartheta}_n-c(1-\hat{\vartheta}_n)\geq 0,\quad \forall n \in\{0,1,\ldots, t-1\}, \\
  0, \quad\text {otherwise}.
  \end{cases}
\end{equation}

\subsection{Quantities of interest}
We are interested in the agent's \textit{quitting} which happens when they stop placing trust: $\{A_t=0\}$. The random variable $\tau$ denotes the number of rounds in which trust was placed until the first `do not place trust' action:
\begin{equation}\label{eq:quitting}
  \tau :=\inf\{t\in\mathbb{N}\cup \{\infty\} : r\hat{\vartheta}_t-c(1-\hat{\vartheta}_t)<0 \}.
\end{equation}
By the definition of $\tau$ in (\ref{eq:quitting}) and $A_t$ in (\ref{eq:Action}), we have
\begin{equation}
A_t = 1,\quad \forall t\leq\tau\:\:\:\text{ and } \:\:\:A_t = 0,\quad\forall t> \tau.
\end{equation}
We thus note that $\hat{\vartheta}_t = \hat{\vartheta}_\tau$ for all $t>\tau$, which arises quite naturally from the model dynamics considering that once the agent has taken the action to not place trust, they also do not get to observe the outcome of the action and thus do not adjust their estimate. This links the agent's actions to the estimate and vice-versa. An institution can have a true trustworthiness $\vartheta$ such that $r\vartheta -c(1-\vartheta)>0$, implying that an agent aware of the true value of $\vartheta$ would place trust forever. It is also possible in such cases that the agent's estimate $\hat{\vartheta}$ continues to adhere to the condition in (\ref{eq:Action}) which means that they never lose trust. In such cases quitting is not a given and a particularly interesting probability to study is that of the event $\{\tau<\infty\}$. We denote this by
\begin{equation}
\pq:=\mathbb{P}(\tau<\infty).
\end{equation}
Furthermore, we are interested in the expected time spent in the system before quitting conditioned on quitting:
\begin{equation}\label{defq}
  q := \mathbb{E}[\tau\,|\, \tau<\infty].
\end{equation}

\begin{figure}
\centering
\subfloat[\centering Single agent model. ]{\includegraphics{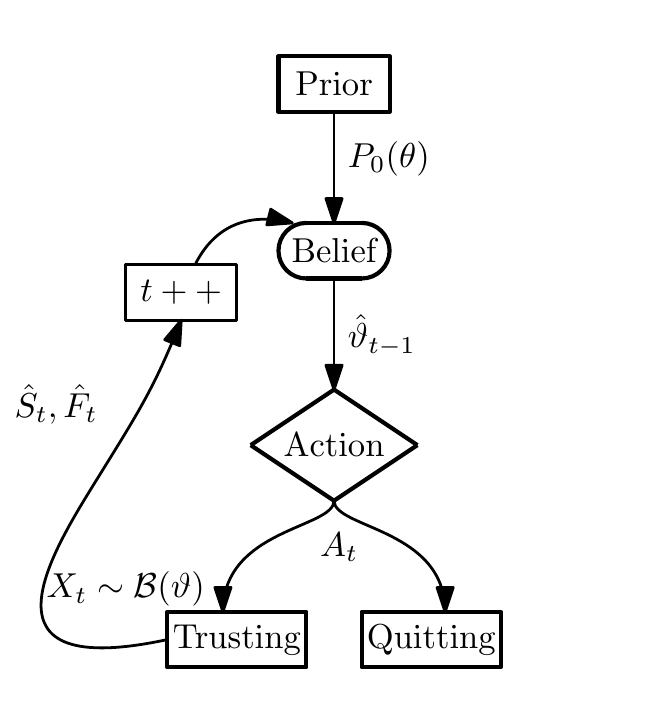}\label{fig:SingleLearnerModel}}
\subfloat[\centering Sample paths from Example~\ref{ex:RW}.]{\includegraphics{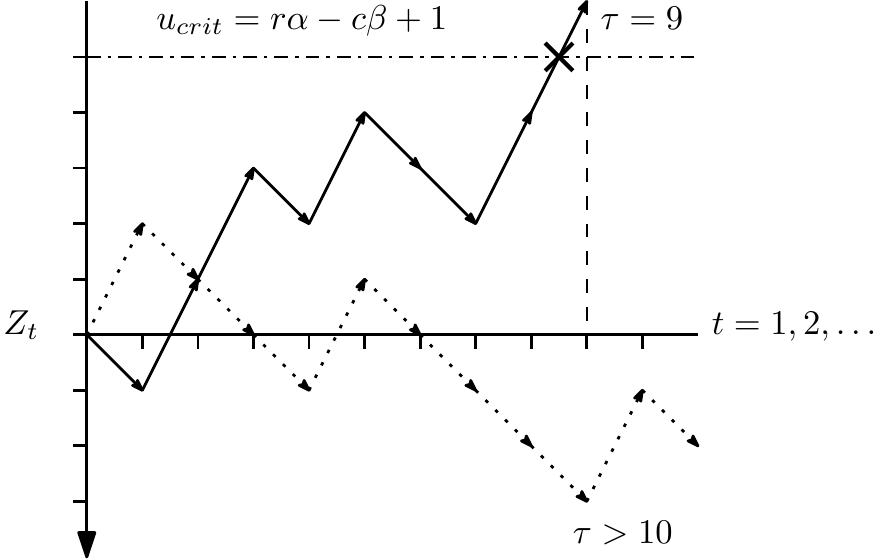}\label{fig:RW}}
\caption{(A) The single agent learner model illustrated conceptually and (B) two example paths of the model dynamics in the random walk interpretation. $\mathcal{B}(\vartheta)$ denoting the Bernoulli distribution with parameter $\vartheta$.}
\end{figure}

We now recapitulate the dynamics with reference to the graphical representation in Figure~\ref{fig:SingleLearnerModel}. The two elements determining the agent's belief distribution at the end of time $t$ are their prior $P_0$ as well as the history of their interaction contained in the random variables $\hat{S}_{t}$ and $\hat{F}_{t}$. Together these result in a trustworthiness estimation of the institution $\hat{\vartheta}_t$ by which the agent makes the choice to either place trust or to quit the trust relationship in round $t+1$. If the agent places trust (thus continuing the process) there is a response from the institution of either honouring the trust (at probability $\vartheta$) or abusing the trust (at probability $1-\vartheta$). Note that all of the model's randomness stems from the response of the institution.

\section{Trusting alone}\label{sec:SinglePlayer}
In this section we analyse the dynamics of the single agent model described in \S\ref{BaseModel}. Our interest lies in the probability of quitting $\pq$ and the expected time to quitting $q$ (conditioned on this occurrence). We consider the dynamics that unfold between a single agent who periodically places (or doesn't place) trust in the institution. We start this section in \S\ref{sec:SingleLearner} with a re-interpretation of the model in terms of a random walk with an absorbing barrier and (potentially asymmetric) step sizes as well as (potentially asymmetric) step probabilities. Subsequently in \S\ref{sec:1LResults} we analyse this random walk model for a number of special cases, focusing on the evaluation of the metrics $\pq$ and $q$.
\subsection{Model re-interpretation}\label{sec:SingleLearner}
We interpret the model described in \S\ref{BaseModel} as a 1-dimensional random walk with step sizes $-r$ and $+c$ at probabilities $\vartheta\in(0,1)$ and $1-\vartheta$ respectively. For an agent holding an initial Beta belief distribution with parameters $\alpha$ and $\beta$ the decision criteria becomes
\begin{equation}
  r\left(\frac{\alpha + \hat{S}_t}{\alpha + \beta + \hat{S}_t + \hat{F}_t}\right) - c\left(1-\frac{\alpha + \hat{S}_t}{\alpha + \beta + \hat{S}_t + \hat{F}_t}\right)\geq 0.
\end{equation}
By rearrangement we find that in order to place trust the agent needs 
\begin{equation}
  c\hat{F}_t -r\hat{S}_t\leq r\alpha-c\beta,
\end{equation} 
which we interpret as a random walk
\begin{equation}
  Z_t:=c\hat{F}_t -r\hat{S}_t,\quad \text{for } t = 0,1,\ldots,
\end{equation}
starting at $Z_0=0$, and having an absorbing barrier at the ``critical'' position 
\begin{equation}\label{eq:ucrit_def}
  \uc(\alpha,\beta, c, r):=r\alpha-c\beta +1.
\end{equation} 
Note that different values of $\alpha$ and $\beta$ correspond to the same behaviour in terms of $p_{\text{quit}}$ and $q_{\text{quit}}$ as long as the value of $u_{\text{crit}}$ is the same. The additional unit in (\ref{eq:ucrit_def}) is due to the decision criteria to place trust including equality (the expected reward must be zero or more) while the formulation of an absorbing barrier describes the first position for which trust is not placed. Note that all the influence of the prior is contained in $u_\text{crit}$. This also allows us to reformulate the probability of quitting as
\begin{equation}
  \pq \equiv \pq(\alpha,\beta, c,r,\vartheta) = \mathbb{P}(\exists t : Z_t\geq \uc),
\end{equation}
the probability of the random walking reaching the absorbing barrier. For ease of reference we refer to $c/(c+r)$ as $\theta_\text{crit}$ because an equivalent condition to $Z_t\geq u_{\text{crit}}$ is that $\hat{\vartheta}<\theta_{\text{crit}}$. For illustrative purposes we provide two sample paths that such a random walk might take in Figure~\ref{fig:RW}.
\begin{example}[Random walk interpretation]\label{ex:RW}
Consider the case with an absorbing barrier at $\uc=5$ arising from parameter values $c=2$, $r=1$, $\alpha = 8$ and $\beta =2$. We interpret this as having a critical trustworthiness $\theta_{\text{crit}}=2/3 = 0.67$. The first such path (depicted as a solid arrows) hits the absorbing barrier at $t=9$, while the second path (depicted as dotted arrows) does not hit the absorbing barrier in the time steps shown and so has $\tau>10$. The values of $\hat{S}_t$ and $Z_t$ are shown in Table~\ref{tab:RW} along with the respective estimated trustworthiness $\hat{\vartheta}_t$.
\begin{table}[ht]
  \centering
  \caption{Sample paths and agent beliefs as random walks.}\label{tab:RW}
  \subfloat[Solid line in Figure~\ref{fig:RW}.]{
  \begin{tabular}{l|cccccccccc}
    $t$ & 0 & 1 & 2 & 3 & 4 & 5 & 6 & 7 & 8 & 9 \\
    \hline
    $\hat{S}_t$ & 0 & 1 & 1 & 1 & 2 & 2 & 3 & 4 & 4 & 4 \\
     $Z_t$ & 0 & -1 & 1 & 3 & 2 & 4 & 3 & 2 & 4 & 6 \\
     $\hat{\vartheta}_t$ & 0.8 & 0.82 & 0.75 & 0.69 & 0.71 & 0.67 & 0.69 & 0.71 & 0.67 & 0.63
  \end{tabular}
  \label{tab:my_label}}
  
  \subfloat[Dotted line in Figure~\ref{fig:RW}.]{
   \begin{tabular}{l|cccccccccccc}
    $t$ & 0 & 1 & 2 & 3 & 4 & 5 & 6 & 7 & 8 & 9 & 10 & 11\\
    \hline
    $\hat{S}_t$ & 0 & 0 & 1 & 2 & 3 & 3 & 4 & 5 & 6 & 7 & 7 & 8 \\
     $Z_t$ & 0 & 2 & 1 & 0 & -1 & 1 & 0 & -1 & -2 & -3 & -1 & -2 \\
     $\hat{\vartheta}_t$ & 0.8 & 0.73 & 0.75 & 0.77 & 0.79 & 0.73 & 0.75 & 0.76 & 0.78 & 0.79 & 0.75 & 0.76
  \end{tabular}}
\end{table}\hfill$\clubsuit$
\end{example}


Regarding the probability of quitting $\pq$, we consider first the case where $\vartheta < c/(c+r)$.
Note that in this case the actual expected utility of placing trust is negative and that the process $Z_t$ has a drift toward the absorbing barrier, meaning that this will be reached eventually with probability 1.
\begin{lemma}[Guaranteed quitting]\label{lem:quit}
If $\vartheta< c/(c+r)$, then $\pq = 1,$ and $\tau<\infty.$
\end{lemma}
The proof (given in Appendix~\ref{AX:proofs}) heavily relies on the law of large numbers. The rest of our investigation in this section takes place in the (more interesting) case where $\vartheta > \theta_\text{crit}$. This case is particularly relevant as it represents cases in which the optimal action for the agent would be to place trust indefinitely yet they do not necessarily do so. The random walk is investigated for general absorbing barriers $u\geq 0$ in order to find recurrence relationships between $p_\text{quit}$ for different values of $u$. We extract the probability of quitting the trust relationship and of the time at which this occurs by using the appropriate $u$ for the parameter values in question.

Considering the model thus far described, an agent either continues to place trust indefinitely and learns the true value of the trustworthiness or quits at some time $t<\infty$. 
\begin{lemma}[Converge or quit]\label{RorC}
A single agent partaking in the trust relationship described with $\vartheta>\theta_{\text{crit}}$ either {\rm (A)} quits at some time $\tau<\infty$, i.e.,
\begin{equation}
  A_t=1,\quad\forall t<\tau, \:\:\:\text{ and } \:\:\:A_t = 0, \quad \forall t\geq \tau,
\end{equation}
or {\rm (B)} continues to place trust indefinitely and has their estimate converge to the true trustworthiness, i.e.,
\begin{equation}
  A_t = 1, \quad \forall t = 1, 2, \ldots,\:\:\:\text{ and }\:\:\: \lim_{t\to \infty}\hat{\vartheta}_t=\vartheta.
\end{equation}
This entails that
\begin{equation}
  \mathbb{P}(\hat{\vartheta} \not\to \vartheta\:\: \&\:\: \tau =\infty) = 0.
\end{equation}
\end{lemma}

The proof follows from the definition of the estimated trustworthiness value and the law of large numbers and is found in Appendix~\ref{AX:proofs}. 

\subsection{Results}\label{sec:1LResults}
In this subsection we study the probability of the agent quitting the trust relationship. We achieve this by finding the absorption probability of the random walk at some level $u$. We define by $\pi(u)$ the probability of hitting an absorbing barrier at $u$ in terms of the distance from the starting point $Z_t=0$ to this absorbing barrier at $u\geq 0$:
\begin{equation}
  \pi(u)\equiv \pi(u,c,r,\vartheta):=\mathbb{P}(\exists t : Z_t\geq u),\quad \forall u\in \mathbb{N}_0.
\end{equation}
Here we suppress the dependence on $c,r$ and $\vartheta$ for ease of reading. Note that $u$ can equal $0$, corresponding to a scenario that absorption is certain. The analysis of $\pi(u)$ is divided into three cases with respect to $r$ and $c$. Two of the cases we characterise analytically, while we present a numerical approximation for the third. Using $u=\uc$ gives the probability of quitting, \textit{i.e.},
\begin{equation}\label{eq:pquc}
\pq(\alpha,\beta,c,r,\vartheta) = \pi(\uc,c,r,\vartheta).
\end{equation}
In \S\S\ref{secp1}-\ref{secc1} we provide the above-mentioned analysis of two cases: 1) $r\in \mathbb{N}$ and $c=1$, and 2) $r=1$ and $c\in \mathbb{N}$. An approximation scheme for the case $r,c\in \mathbb{N}$ is presented in Appendix~\ref{sec:pncn}. The split is a result of the fact that it is not possible to find a closed form result for the general case. The techniques used in the two cases presented also differ substantially and should therefore be viewed separately.

\subsubsection{Case \texorpdfstring{$c=1$}{c=1}, and \texorpdfstring{$r\in\mathbb{N}$}{q in N}}\label{secp1}
In this case the random walk exhibits a useful memorylessness property. To see this, observe that the walk can only go up levels one at a time, while on its way down it can skip levels. We therefore first define the probability of the random walking climbing by one level for some $t\in\mathbb{N}$:
\begin{equation}\label{rhodef}
  \varrho :=\mathbb{P}(\exists t : Z_t\geq 1).
\end{equation}
As a consequence of the strong Markov property, we note that the dynamics after attaining 1 level is independent of the history by which this was done. This means that the probability of going up $u$ levels is simply the probability of $u$ times sequentially going up 1 level, so that $\pi(u)=\varrho^u$. A formal expression of the probability of quitting, including a characterization of $\varrho$, is presented in the following lemma which is proved in Appendix~\ref{AX:proofs}.
\begin{lemma}[Quitting probability when $c=1$, and $r\in\mathbb{N}$]\label{absp1}
Suppose $\vartheta \geq 1/(1+r)$. The probability of the corresponding random walk with parameter values $c=1$, $r\in\mathbb{N}$ reaching an absorbing barrier at $u$ satisfies
\begin{equation}\label{eq:pip1qN}
  \pi(u,1,r,\vartheta)=\varrho(\vartheta)^u,\quad \forall u\in\mathbb{N}_0,
\end{equation}
in which $\varrho(\vartheta)=\pi(1,1,r,\vartheta)$ is the unique solution in the range $[0,1)$ to the equation 
\begin{equation}\label{eq:rhosol}
  \varrho(\vartheta) = (1-\vartheta) + \vartheta \varrho(\vartheta)^{r+1}.
\end{equation}
To find the probability of quitting we use $\uc = r\alpha-\beta +1$ in 
\begin{equation}\label{eq:pqr1}
  \pq(\alpha,\beta, 1, r, \vartheta) = \pi(\uc,1,r, \vartheta). 
\end{equation}
\end{lemma}

\begin{figure}
  \centering
  \input{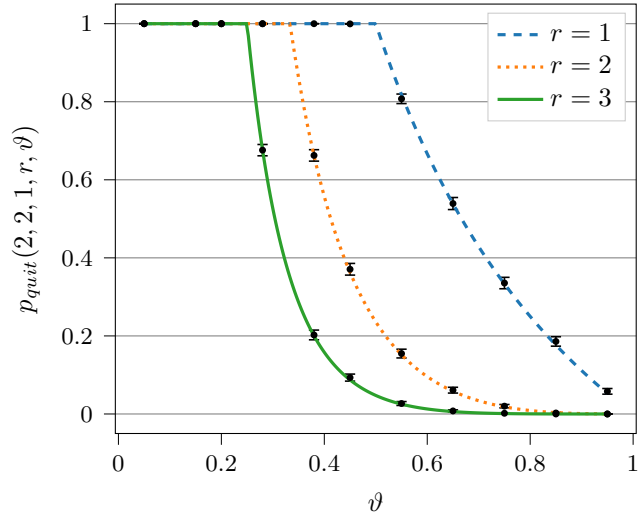}
  \caption{The probability of a single agent quitting $p_\text{quit}(\alpha,\beta,c,r,\vartheta)$ plotted against $\vartheta$ for different values of $r$ while $\alpha = \beta = 2$ and $c=1$. Analytical results are plotted as lines, while simulated results (4\,000 iterations) are shown as points with 95\% confidence intervals.}
  \label{fig:Ruinp1}
\end{figure}

In Figure~\ref{fig:Ruinp1} we plot numerical results of the quitting probability denoted $\pq(\alpha,\beta, c, r,\vartheta) = \pi(\uc,c,r,\vartheta)$ for the subcases $r=1,2,3$. The Beta prior belief distribution is given by shape parameters $\alpha = \beta = 2$. The lines are theoretical results obtained by Lemma~\ref{absp1}, and the dots (with confidence intervals) correspond to simulated results which corroborate the analytical results. The shape of the quitting probability is explicable by noting that a lower trustworthiness $\vartheta$ consistently leads to a higher quitting probability. The results of Lemmas~\ref{lem:quit} and~\ref{absp1} are illustrated as the probability of quitting is unity where $\vartheta<\theta_\text{crit}$ and follow (\ref{eq:pqr1}) thereafter. 

We now turn our interest to the expected time an agent spends in the system. Note that the time of quitting $\tau$ in terms of the random walk notation is such that
\begin{equation}
  \tau(u)\equiv \tau(u,c,r,\vartheta)=\inf\{t\in\mathbb{N}_0:Z_t\geq u\}.
\end{equation} Then we are interested in the expectation of $\tau(u)$ conditioned on $\mathbf{1}\{\tau(u)<\infty\}$ which we now know occurs with probability $\pi(u)$. Specifically we investigate the quantity
\begin{equation}
  \mathbb{E}[\tau(u) \,|\,\tau(u)<\infty]=\frac{\mathbb{E}\left[\tau(u) \mathbf{1}\{\tau(u)<\infty\}\right]}{\pi(u)}, \text{ for }u\in\mathbb{N}_0. 
\end{equation}
We use the techniques from probability generating functions in order to describe the expectation of the distribution at hand. As such we inspect a number related to the expected time to reach the first level, $\tau(1)$:
\begin{equation}
  \varphi(z):= \mathbb{E}[z^{\tau(1)}\mathbf{1}\{\tau(1)<\infty\}], \quad z\in [-1,1],
\end{equation}
and note that 
\begin{equation}
  \mathbb{E}[z^{\tau(u)}\mathbf{1}\{\tau(u)<\infty\}] = \varphi(z)^u, \quad\forall u\in\mathbb{N}_0.
\end{equation}
This (again) follows from the fact that all levels $\{1,\ldots,u-1\}$ need to be passed first in order reach $u$. Thus, the expected time to reach the $u^{\text{th}}$ level is simply $u$ times the expected time to reach the first level.

\begin{lemma}[Expected time to quitting when $c=1$, and $r\in\mathbb{N}$]\label{lem:exTqN}
If $r\in\mathbb{N}$, $c=1$ and $\vartheta \geq 1/(r+1)$ then for all $u\in \mathbb{N}$, $\tau(u)$ satisfies
\begin{equation}
  \mathbb{E}[\tau(u)\,|\, \tau(u)<\infty]= u\frac{\varphi'(1)}{\varrho},\quad \forall u\in\mathbb{N},
\end{equation}
where $\varphi'(z)$ is the derivative from below of the function $\varphi(z)$ which, for any given $z\leq|1|$, is the unique solution to 
\begin{equation}
  \varphi(z) = z(1-\vartheta)+z\vartheta\,\varphi(z)^{r+1},
\end{equation}
and where $\varrho$ is the unique solution to equation (\ref{eq:rhosol}). In particular,
\begin{equation}\label{eq:taup1qN}
   \mathbb{E}[\tau(u)\,|\, \tau(u)<\infty] =\frac{u\varrho}{(1-\vartheta)(r+1)-r\varrho},\quad \forall u\in\mathbb{N}.
\end{equation}
To find the expected quitting time (conditional on quitting), we use $\uc = r\alpha-\beta +1$ in 
\begin{equation}\label{eq:pqc1}
  q(\alpha,\beta,1,r,\vartheta)=\tau(\uc,1,r,\vartheta). 
\end{equation}
\end{lemma}
To find the expected time in the system one starts by solving (\ref{eq:rhosol}) with the appropriate value of $r$ for the unique solution of $\varrho\in (0,1)$. This in turn is substituted into (\ref{eq:taup1qN}) along with $u=\uc$, resulting in the desired expression for $\tau(\uc,1,r,\vartheta)$. For details, see the proof in Appendix~\ref{AX:proofs}.

\begin{figure}
  \centering
  \input{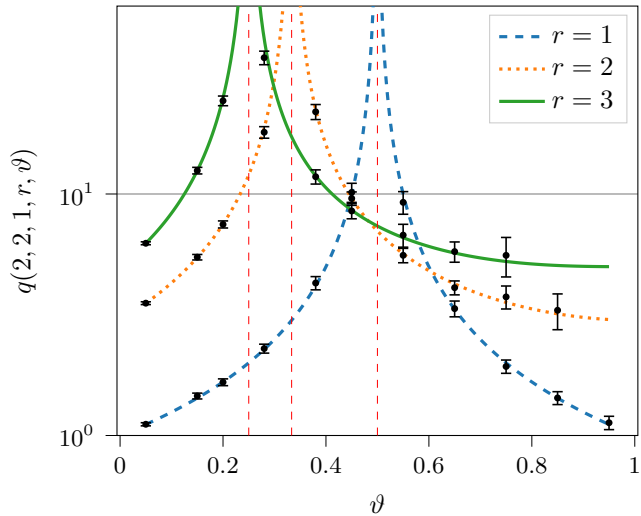}
  \caption{Expected time in the system conditioned on the agent quitting at some $t<\infty$ plotted against $\vartheta$ for different values of $r$ while $\alpha=\beta = 2$ and $c=1$ on a log-linear axis. Analytical results are plotted as lines, while simulated results are shown as points with 95\% confidence intervals.}
  \label{fig:ExTp1}
\end{figure}

We plot numerical results as well as simulated results of the expected time in the system conditioned on the agent's eventual quitting $q(\alpha,\beta,1,r,\vartheta)$ in Figure~\ref{fig:ExTp1} for the subcases $r=1,2,3$ on a log-linear axis. The Beta prior belief distribution 
is characterised by shape parameters $\alpha = \beta = 2$. We notice that the expected time is increasing toward an asymptote positioned on $\vartheta = 1/(1+r) = \theta_\text{crit}$. The position of this asymptote is due to the property of the model: quitting is still guaranteed for $\vartheta = \theta_\text{crit}$ but the time it takes has an infinite expectation ({\it cf}.\ the symmetric Bernoulli walk). This initial increase makes sense, as the probability of making a detour away from quitting is increasing with $\vartheta$ though it cannot escape forever in this region. The subsequent decrease as $\vartheta> \theta_{\text{crit}}$ results from the fact that, as $\vartheta$ increases, if quitting takes place then it does so earlier. This is because more time in the relationship exposes the agent to more unbiased information which is likely to indicate that the institution is trustworthy.

\subsubsection{Case \texorpdfstring{$r=1$}{r = 1}, and \texorpdfstring{$c\in\mathbb{N}$}{c in N}}\label{secc1}
In this case we notice that the walk $Z_t$ loses its memoryless property: steps upward now have size $c\in\mathbb{N}$ and so it is possible that levels are skipped along the way. This means that a relationship of the type $\pi(u)=\varrho^u$ no longer holds. 
We thus see that one has no guarantee of which levels are reached on the way to quitting, and therefore one cannot use the approach that we relied upon in the $c=1$ case. We nevertheless obtain the following result. In this case the appropriate value of $u$ is given by $u_\text{crit}=\alpha -c\beta+1$.

\begin{lemma}[Quitting probability when $r=1$, and $c\in\mathbb{N}$]\label{lemq1}
If $r=1$, $c\in\mathbb{N}$ and $\vartheta \geq c/(c+1)$, then the probability of quitting is given 
\begin{equation}
  \pq = \pi(\uc,c,1,\vartheta),
\end{equation}
where $\pi(u,c,1,\vartheta)$ satisfies
\begin{equation}
  \pi(u)=\frac{\xi^{(u)}(0)}{u!}, \quad \forall u\in\mathbb{N};
\end{equation}
here $\xi^{(u)}(0)$ is the $u^{\text{th}}$ derivative of the function $\xi(w)$ in $w=0$, where $\xi(w)$ is defined by
\begin{equation}
  \xi(w)
  =\frac{[(1-\vartheta)(c+1)-1]w^2+[2-(1-\vartheta)(c+1)]w-(1-\vartheta) w^{c+1}-\vartheta}{(1-\vartheta) w^{c+2} -(1-\vartheta) w^{c+1} -w^2+(1+\vartheta)w-\vartheta}, \:\:\forall w\in(-1,1).
\end{equation}\end{lemma}

The proof (in Appendix~\ref{AX:proofs}) uses a translation between the maximum of our random walk on one hand, and the minimum of a random walk with $c=1$ and $r\in\mathbb{N}$ on the other hand. 

We observe that $\xi(u)$ is of the form ${P(w)}/{Q(w)}$, where $P(w)$ and $Q(w)$ are polynomials of degree $c+1$ and $c+2$ respectively. We evaluate values for $\pi(u)$ by differentiating both sides of $Q(w)\,\xi(w)=P(w)$, $u$ times and substituting $w=0$ in order to get the expression for $\pi(u)$.
Here $\xi(w)$ can be interpreted as a probability generating function with coefficients $a_u = \pi(u)$, which allows calculation of the ruin probabilities by differentiating $u$ times, setting $w=0$ and dividing by $u!$, a standard procedure from the theory of probability generating functions.

\begin{figure}[h!]
  \centering
  \input{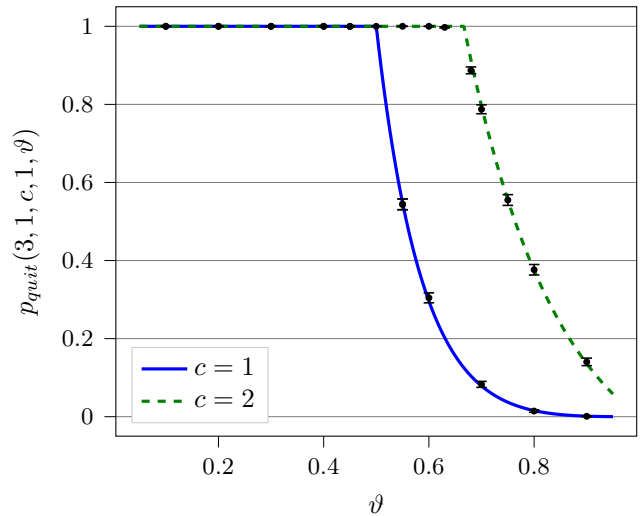}
  \caption{The probability of a single agent quitting plotted against $\vartheta$ for different values of $c$ while $\alpha = 3, \beta = 1$ and $r=1$. Analytical results are plotted as lines, while simulated results (4\,000 iterations) are shown as points with 95\% confidence intervals.}
  \label{fig:ruinq1}
\end{figure}

The resulting quitting probabilities are plotted in Figure~\ref{fig:ruinq1} for the first two values of $c$. As expected, quitting remains certain until $\vartheta>c/(c+1)$, the stability condition. The prior belief parameters were chosen in order to ensure that the agent does not quit in round 0. The effect of the prior belief parameters is that $u_\text{crit}=3$ for $c=1$ and $u_\text{crit}=2$ for $c=2$.

As before, we are also interested in the expected time that an agent spends in the system placing trust, conditional on them quitting eventually. Similar hurdles arise here as a result of the potentially larger step in the direction of the barrier, though in this analysis these are handled without much extra machinery. The following lemma expresses the expected time that an agent places trust until quitting conditioned on their quitting eventually. 

\begin{lemma}[Expected time to quitting when $r=1$, and $c\in\mathbb{N}$]\label{lem:exTpN} Define
\begin{equation}
  \Phi(w,z) :=\frac{(1-\vartheta)z\left(\frac{w-w^{c+1}}{1-w}\right)-\vartheta z \,\bar\varphi(z)}{1-(\vartheta z)/w -w^c z (1-\vartheta)}, \quad w\in(-1,1),
\end{equation}
and let $w(z)$, for any given $z\in(-1,1)$, denote the unique solution for $w\in(-1,1)$ in
\begin{equation}
w^{c+1} (1-\vartheta)z+\vartheta z= w,
\end{equation}
and
\begin{equation}\bar\varphi(z) := \frac{(1-\vartheta)zw(z)+\vartheta z-w(z)}{1-w(z)}.\end{equation}
Suppose $\vartheta > c/(c+1)$ and the utility parameters of the agent are given by $c\in\mathbb{N}$ and $r=1$. The expected time for the corresponding random walk to hit an absorbing barrier at $u$ satisfies
\begin{equation}\label{eq:Etau}
 \mathbb{E}[\tau(u) \,|\,\tau(u)<\infty] = \frac{\varphi'(1,u)}{\varphi(1,u)},
\end{equation}
where
\begin{equation}
  \varphi(z,u):= \left.\frac{\partial^u}{\partial w^u}\frac{\Phi(w,z)}{u!}\right|_{w=0}.
  \end{equation}
\end{lemma}
In order to fully express $\Phi$ in terms of only $w,\vartheta,$ and $z$, one first uses the fact that a root of the denominator is necessarily also a root of the numerator (to keep $\Phi(w,z)$ bounded), and using it to express $\varphi(z,1)$ in terms of $z$. The proof, presented in Appendix~\ref{AX:proofs}, makes use of the object $\varphi(z,u):=\mathbb{E}[z^{\tau(u)}\mathbf{1}\{\tau(u)<\infty\}]$ which is then used in the generating function 
\begin{equation}\Phi(w,z):=\sum_{u=1}^\infty w^u\varphi(z,u)\end{equation} for $w\in(-1,1)$ from which we obtain the result. 

\begin{figure}[htb]
  \centering
  \input{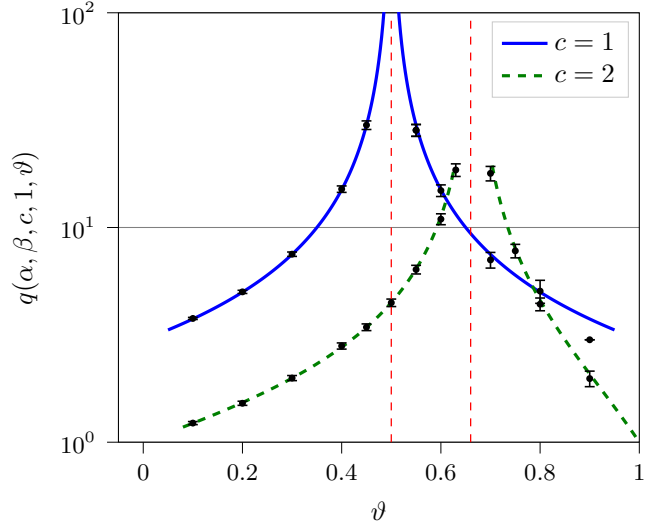}
  \caption{Expected time in the system conditioned on the agent quitting at some $t<\infty$ plotted against $\vartheta$ for different values of $c$ while $\alpha = 3, \beta = 1$ and $r=1$. Analytical results are plotted as lines, while simulated results are shown as points with 95\% confidence intervals.}
  \label{fig:TauR1}
\end{figure}

In Figure~\ref{fig:TauR1} we plot numerical results as well as simulated results of the expected time in the system conditioned on the agent's eventual quitting $q(\alpha,\beta,c,1,\vartheta)$. We do so for the instances $c=1,2$ using the Beta prior belief distribution determined by shape parameters $\alpha = 3$ and $\beta = 1$, {\it i.e.}, the same parameter choices as the ones used in Figure~\ref{fig:ruinq1}. As in the case with $c=1$ and $r\in \mathbb{N}$, we notice that the expected time is increasing toward an asymptote, which is now positioned at $\vartheta = c/(c+1)$. After this critical $\vartheta$ there is a decrease as failing becomes more centered around the early time steps. More time in the system implies that the agent is exposed to more unbiased experience with the institution, and therefore they have a greater chance of remaining in the trust relationship.

\section{Trusting together}\label{sec:2Player}
In this section we extend the model from one truster agent to two truster agents. To this end, we enhance the model with a communication mechanism (between the truster agents, that is) that retains the assumptions of myopically rationality.

We consider agents who communicate as frequently with one another as they make decisions to place or not place trust. The nature of this interaction is considered in two forms. The first is that the agents share completely their own interaction history with each other. The second form is that the agents do not communicate about their outcomes at all, and only observe which action is being taken by the other agent in each round.

\subsection{Shared modelling aspects}
In both mechanisms we have two agents, each with their own interaction with the institution. As such, we have decisions for each agent per round from the action set $\mathcal{A}=\{0,1\}^2$ in which $A_{i,t}=1$ if agent $i\in\{1,2\}$ places trust in round $t$. For each round $t$ and agent $i$,
\begin{equation}
  X_{i,t} =\begin{cases}
  1, \text{ with probability }\vartheta \\
  0, \text{ with probability }(1-\vartheta),
  \end{cases}
\end{equation}
where $X_{i,t}=1$ indicates that the trust that was placed in that round was subsequently honoured by the institution. Similarly to the single learner case the agents use
\begin{equation}
  \hat{S}_{i,t}=\sum_{n=0}^tX_{i,n}\mathbf{1}_{\{A_{i,n}=1\}},
\end{equation}
to keep track of how many times trust was honoured to them in rounds when they placed trust. We also define the number of times trust was abused to each agent $i = 1,2$ by
\begin{equation}
  \hat{F}_{i,t}=\sum_{n=0}^t(1-X_{i,n})\mathbf{1}_{\{A_{i,n}=1\}}.
\end{equation}
The agents are equipped with the same Beta prior belief distribution as before.
The prior belief distribution (with shape parameters $\alpha_{i},\beta_{i}\in\mathbb{N}$) for agent $i$ is therefore
\begin{equation}
  P_{i,0}=B(\theta;\alpha_{i},\beta_{i})= \frac{\theta^{\alpha_{i}-1}(1-\theta)^{\beta_{i}-1}}{\int_0^1 y^{\alpha_{i}-1}(1-y)^{\beta_{i}-1}\dY}, \quad \theta \in [0,1].
\end{equation}
For simplicity, from now on we consider only the case of homogeneous priors ({\it i.e.}, $\alpha_1 = \alpha_2$ and $\beta_1=\beta_2$). It should be noted that one can deal with non-homogeneous priors in the exact same way, where one should specify what each of the agents knows about their neighbour's prior. One could assume that they know it exactly, or hold some belief distribution on it; evidently, the formulation becomes more tedious but is conceptually straightforward. 

As we are only considering the homogeneous case, we drop the subscript on the prior belief distribution, calling it simply $P_0(\cdot)$.
Let the information received by agent $i$ in round $t$ be denoted $I_{i,t}(\theta)$. The precise formulation of this information is a modelling outcome of the choice of communication; see (\ref{eq:OR_info}) of \S\ref{info} for the first mechanism, and (\ref{eq:OA_info}) of \S\ref{obs} for the second mechanism. The resulting belief update rule becomes
\begin{equation}\label{eq:gen2pl}
  P_{i,t}(\theta)= \frac{\theta^{\hat{S}_{i,t}}(1-\theta)^{\hat{F}_{i,t}}P_{0}(\theta)I_{i,t}(\theta)}{\int_0^1 y^{\hat{S}_{i,t}}(1-y)^{\hat{F}_{i,t}}P_{0}(y)I_{i,t}(y)\,\dY}, \quad \theta \in [0,1].
\end{equation}
The difference between (\ref{eq:gen2pl}) and (\ref{eq:belief_1pl}) lies in the presence of the information received by the agent. Subsequently the agents use the mean of their belief distribution as a point estimate for the true trustworthiness of the institution and use this to make the decision whether or not to place trust in round $t+1$. If we define
\begin{equation}
  \hat{\vartheta}_{i,t} = \int_0^1\theta P_{i,t}(\theta)\,\dT, \quad \theta \in [0,1],
\end{equation}
\textit{i.e.}, the mean belief held by agent $i$ at the end of round $t$, then agent $i$ takes the action $A_{i,t+1}$ in round $t+1$ where
\begin{equation}
A_{i,t}=\begin{cases}\label{eq:multiagentcondition}
1, \text{ if }r\hat{\vartheta}_{i,n}-c(1-\hat{\vartheta}_{i,n})\geq 0,\quad \forall n \in \{0,1,\ldots,t-1\},\\
0, \text{ otherwise.}
\end{cases}
\end{equation}
Note that the condition in (\ref{eq:multiagentcondition}) can be rewritten as $\hat{\vartheta}_{i,n}\geq \theta_{\text{crit}} ,\:\forall n\in \{0,1,\ldots,t-1\}$, with $\theta_{\text{crit}}:=c/(c+r)$.
In the next two subsections we discuss the two different communication mechanisms in more detail. 
\subsection{Observable rewards (OR)}\label{info}
\begin{figure}
  \centering
  \includegraphics{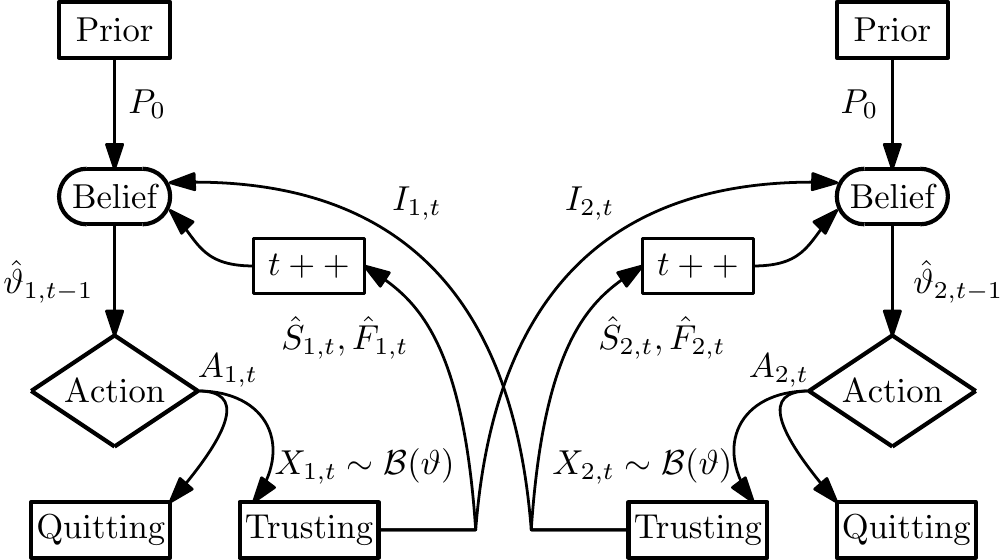}
  \caption{The observable rewards model of learning.}
  \label{fig:InfoMod}
\end{figure}

This case considers agents that communicate fully with one another about their experiences with the institution. The dynamics of this model are represented graphically in Figure~\ref{fig:InfoMod}. The extension from the single learner model (represented in Figure~\ref{fig:SingleLearnerModel} for comparison) to this model consists of the shared information between agents which entails the response of the institution to trust that was placed.

Each agent thus has same information available to them, and under homogeneous prior beliefs they have the same estimate of the true trustworthiness. In this case we define the information agent $i$ receives from their neighbour $j\neq i$ in round $t$,
\begin{equation}\label{eq:OR_info}
  \theta \mapsto I_{i,t}:=\theta^{\hat{S}_{j,t}}(1-\theta)^{\hat{F}_{j,t}}, \quad \theta \in [0,1].
\end{equation}
We see that this information contains for each agent exactly the interaction history of the other agent, hence we have the line from agent $i$'s ``Trusting'' action to agent $j$'s ``Belief'' distribution in Figure~\ref{fig:InfoMod}. 

As mentioned, the agents' estimated values of the true trustworthiness are the same. Thus we can drop the subscript $i$ from $\hat{\vartheta}_{i,t}$ and note that it is given by
\begin{equation}
  \hat{\vartheta}_t=\frac{\alpha+\sum_{i=1}^2\hat{S}_{i,t}}{\alpha+\beta+\sum_{i=1}^2\hat{S}_{i,t}+\sum_{i=1}^2\hat{F}_{i,t}}.
\end{equation}
Defining $Z_t=c\sum_{i=1}^2\hat{F}_{i,t}-r\sum_{i=1}^2\hat{S}_{i,t}$, similarly as in the single-agent model we get a random walk that takes steps
\begin{equation}\label{eq:OR}
  Z_{t+1}=\begin{cases}
  Z_t + 2c, \quad &\text{with probability }(1-\vartheta)^2\\
  Z_t + c-r, &\text{with probability }2\vartheta(1-\vartheta)\\
  Z_t -2r, &\text{with probability }\vartheta^2,\\
  \end{cases}
\end{equation}
with an absorbing barrier at $u=r\alpha-c\beta+1$.

In the case of heterogeneous priors the formulation of this model does not change dramatically. The fates of the agents would not be tied as before and so there would simply be two random walks defined as in (\ref{eq:OR}), one for each agent. Each random walk would have its own absorbing barrier defined by $u_i=r\alpha_i-c\beta_i+1$. Recall that $\alpha_i,\beta_i,r$ and $c$ for all $i$ are taken from $\mathbb{N}$ and so $u_i$ too is an integer for both $i$. If at some time $t_q$ one of the agents quit, then the remaining agent simply continues according to the dynamics of the single agent model with the new transitions defined by
\begin{equation}
  Z_{t+1} = \begin{cases}
  Z_t + c, \quad &\text{with probability } (1-\vartheta)\\
  Z_t -r, &\text{with probability }\vartheta,
  \end{cases}
\end{equation}
for $t>t_q$.
\subsection{Observable actions (OA)}\label{obs}
In this setting the agents do not communicate explicitly by sharing information about their interaction history. Instead, they observe the \emph{actions} of their neighbouring agent, and use this to infer something about the possible histories their neighbouring agent may have experienced leading to such a result. This is where subtle intricacies arise, as rational agents need to keep in mind that their own actions are observed by their neighbour and thus will also affect the decisions made by their neighbour.

\begin{figure}[ht]
  \centering
  \includegraphics{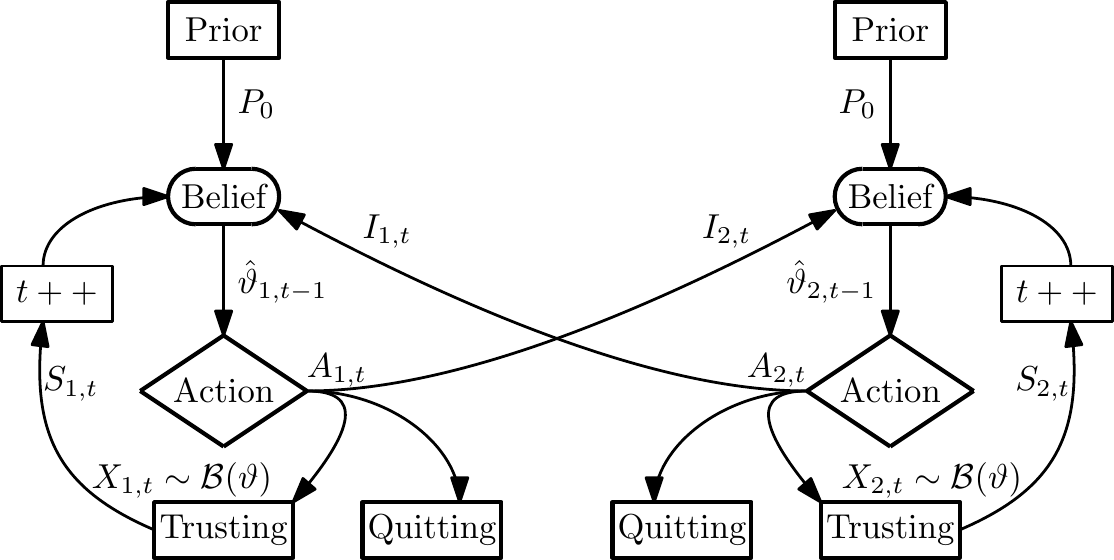}
  \caption{The observable actions model of learning.}
  \label{fig:ObsNet}
\end{figure}

The extent of the communication between the two agents in this model is binary indicating whether or not trust was placed at the end of round $t$. This information is only incorporated into the agents' belief distributions for round $t+1$. This model is depicted graphically in Figure~\ref{fig:ObsNet}. The crux of this model is the line from one agents ``Action'' decision to the other agent's ``Belief'' distribution. The action of not placing trust implies that the agent exits the system forever. The quitting agent does not continue to update their belief based on their neighbour's actions. The information sent by the quitting agent is used as the information received by the not-quitting agent for the remainder of the not-quitting agent's tenure in the system.

The nature of the communication necessitates separate definitions for the information sent with a ``yes'' signal, when the agent places trust, and for the information sent with a ``no'' signal, when the agent does not place trust. As such we define for $i=1,2$, $j\neq i$ and $t\in \mathbb{N}$,
\begin{equation}\label{eq:InfoDefBase}
I_{i,t}:= 
\begin{cases}
    Y_{i,t}\quad &\text{if }t<\tau_j,\\
    N_{i,\tau_j}&\text{if }t\geq\tau_j,\\
\end{cases} 
\end{equation}
where $\tau_j$ is the round in which agent $j$ quits for $j=1,2$. We formally define $Y_{i,t}$ and $N_{i,t}$ in (\ref{eq:OA_info}) and (\ref{eq:NoSig}) respectively. By the rationality assumption, the agents use the binary information along with the knowledge that their own outbound information up until round $t-1$ was used rationally by the other agent. With this they infer a range of possible histories that may have lead to the decision taken by their neighbour. 

\begin{figure}
\centering
\subfloat[\centering $w_x(t)$ counted for a ``yes'' signal.]{\includegraphics[scale=0.85]{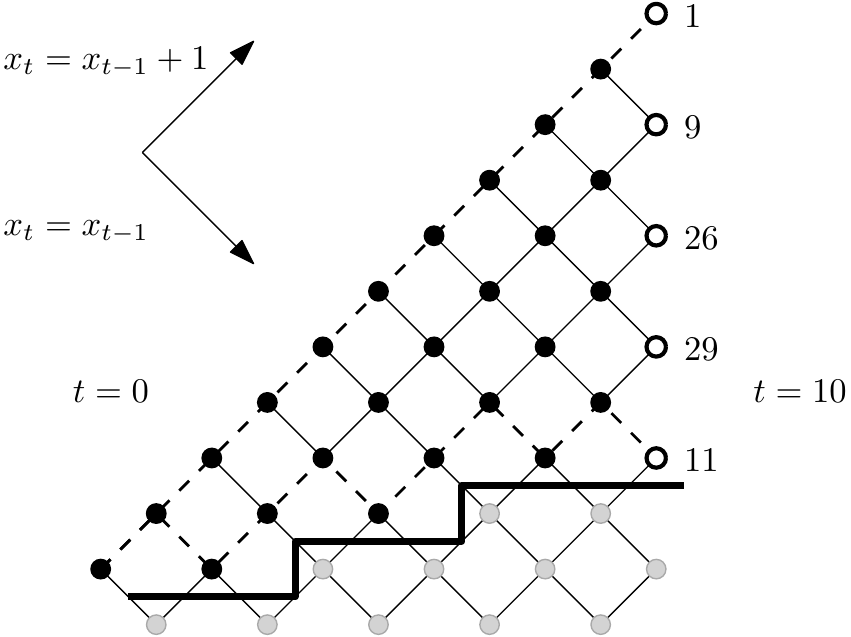}\label{fig:w}}
\subfloat[\centering $w_x(t)$ counted for a ``no'' signal.]{\includegraphics[scale=0.85]{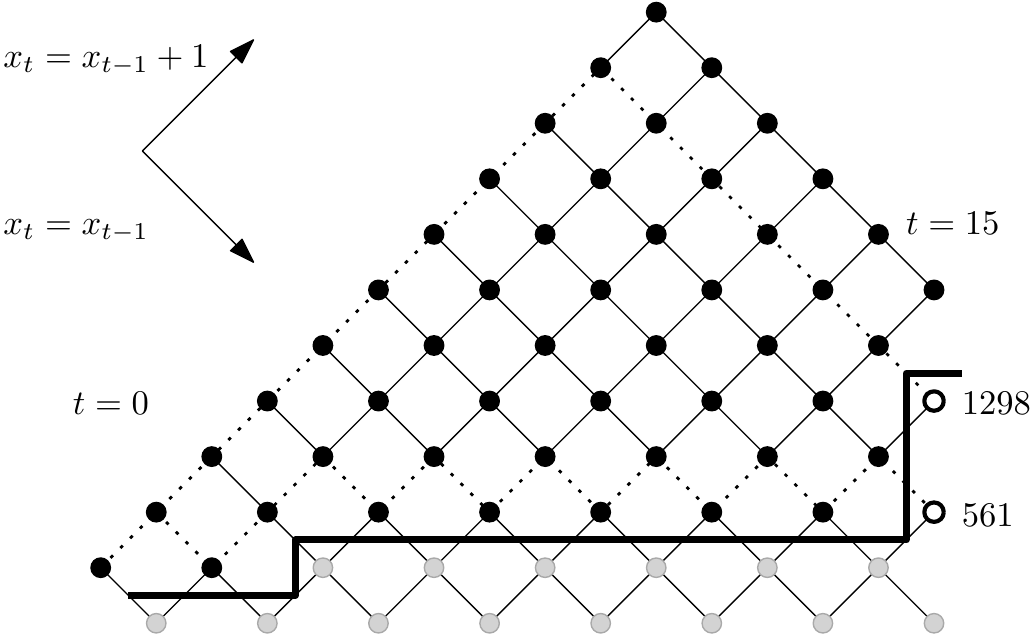}\label{fig:wQuit}}
\caption{An illustration of the weighting $w_x(t)$ used in the interpretation of the observed action.}
\end{figure}

We define the belief distributions $R_t^i$, formed by taking only the report received by agent $i$ (signal sent at time $t$ by agent $j$ to agent $i$) into account:
\begin{equation}
  R_t^i = \frac{P_{0}(\theta) I_{i,t}(\theta)}{\int_0^1P_{0}(y) I_{i,t}(y)\,\dY},\quad\theta \in [0,1].
\end{equation}
We then define the auxiliary belief distribution resulting from a combination of $R_t^i$ and $\theta^x(1-\theta)^{t-x}$ where $x$ is one of the possible number of positive experiences giving $\mathcal{D}_t(x)$ the distribution:
\begin{equation}
  \mathcal{D}_t(x) = \frac{\theta^x(1-\theta)^{t-x}R_{t}^i}{\int_0^1 y^x(1-y)^{t-x}R_{t}^i\,\dY},\quad\theta \in [0,1].
\end{equation}
This allows us to define the set of permitted positions of agent $i$, $\mathcal{J}^i_t$ as
\begin{equation}
  \mathcal{J}_t^i = \big\{x\in \{0,1,\ldots ,t\} \,|\,\mathbb{E}_{\theta\sim \mathcal{D}_{t}(x)}[\theta] \geq \theta_\text{crit}\big\}.
\end{equation}
We have that $\mathcal{D}_{t}(x)$ for all $x\in \mathcal{J}^i_t$ are the possible belief distributions held by agent $i$. These possible belief distributions are then summed in $I_{j,t}$ weighted according $w_x(t-1)$ the number of ways in which these $x$ positive experiences may have been realised (\textit{i.e.}, the number of histories resulting in $x$ positive experiences by time $t$). Defining $w_x(t)$ is done recursively:
\begin{equation}
  w_x(t) =\begin{cases}
  w_x(t-1) + w_{x-1}(t-1) \quad &\text{ if $(x-1)\in \mathcal{J}_{t-1}$}\\
  w_x(t-1) & \text{ if $(x-1)\notin \mathcal{J}_{t-1}$}\\
  1 & \text{ if $x=t$}\\
  0 &\text{ if $x>t$ or $x\notin \mathcal{J}_{t-1}$},\\
  \end{cases} \text{for }x \in [0,t]\text{ and }t\in \mathbb{N}
\end{equation}
with initial condition $w_x(1)=1$ for all $x\in \mathcal{J}_1$. The information received by agent $i$ from agent $j$ considering that agent $j$'s decision was to place trust in round $t$ is thus defined as
\begin{equation}\label{eq:OA_info}
  \theta \mapsto Y_{i,t}:=\sum_{x\in \mathcal{J}_{t-1}^j}w_x(t-1) \,\theta^x(1-\theta)^{t-1-x}, \quad \theta \in [0,1].
\end{equation}
We can interpret $w_x(t)$ as the number of walks from $(0,0)$ to $(x,t-x)$ that retain $x_\tau\in\mathcal{J}_\tau$ for $\tau = 1,2,\ldots,t$. Figure~\ref{fig:w} illustrates the permitted walks for a ``yes'' signal with the boundary between $x$ in and out of $\mathcal{J}_t$ being represented as the thick horizontal line.

In order to define the information sent with a ``no'' signal, we define the range of values which would result in ``yes'' signals up until round $t-1$ and a ``no'' in round $t$:
\begin{equation}
  \mathcal{K}_t^i=[\max\{0,\inf_x \mathcal{J}_{t-1}^i\},\ldots, \inf_x \mathcal{J}_t^i-1].
\end{equation}
Subsequently, we have the information sent in a ``no'' signal:
\begin{equation}\label{eq:NoSig}
\theta \mapsto N_{i,t} := \sum_{x\in\mathcal{K}_t} w_x(t)\, \theta ^x (1-\theta)^{t-1-x}, \quad\theta \in [0,1]
\end{equation}
where $w_x(t)$ is defined as above. The number of permitted walks to a ``no'' signal at time $t$ is simply the number of permitted walks to a ``yes'' signal at time $t-1$ which are also one negative experience away from quitting. This is illustrated in Figure~\ref{fig:wQuit} with a horizontal line indicating the boundary between $x$ in and out of $\mathcal{J}_t$ and the dotted line indicating the boundary between $x$ in and out of $\mathcal{K}_t$. Note that the information sent by a ``no'' signal in subsequent rounds (after the first such signal) does not change, as we allude to in (\ref{eq:InfoDefBase}).

The agents finally construct an estimate by combining their own information with  the signal information:
\begin{equation}
  P_{i,t}(\theta) = \frac{\theta^{\hat{S}_{i,t}}(1-\theta)^{\hat{F}_{i,t}}P_0(\theta)I_{j,t}(\theta)}{\int_0^1y^{\hat{S}_{i,t}}(1-y)^{\hat{F}_{i,t}}P_0(y)I_{j,t}(y)\,\dY}, \quad\theta \in [0,1],
\end{equation}
and place trust in round $t+1$ if $\mathbb{E}_{\theta\sim P_{i,t}(\theta) }[\theta]\geq \theta_{\rm crit}$.

\subsection{Illustration of the differences between the models presented}\label{sec:difModels}
In this subsection we present a set of example experiences for two players. These serve to demonstrate the workings of the dynamics in the OA model and the OR model. 

\begin{figure}
  \centering
  \includegraphics{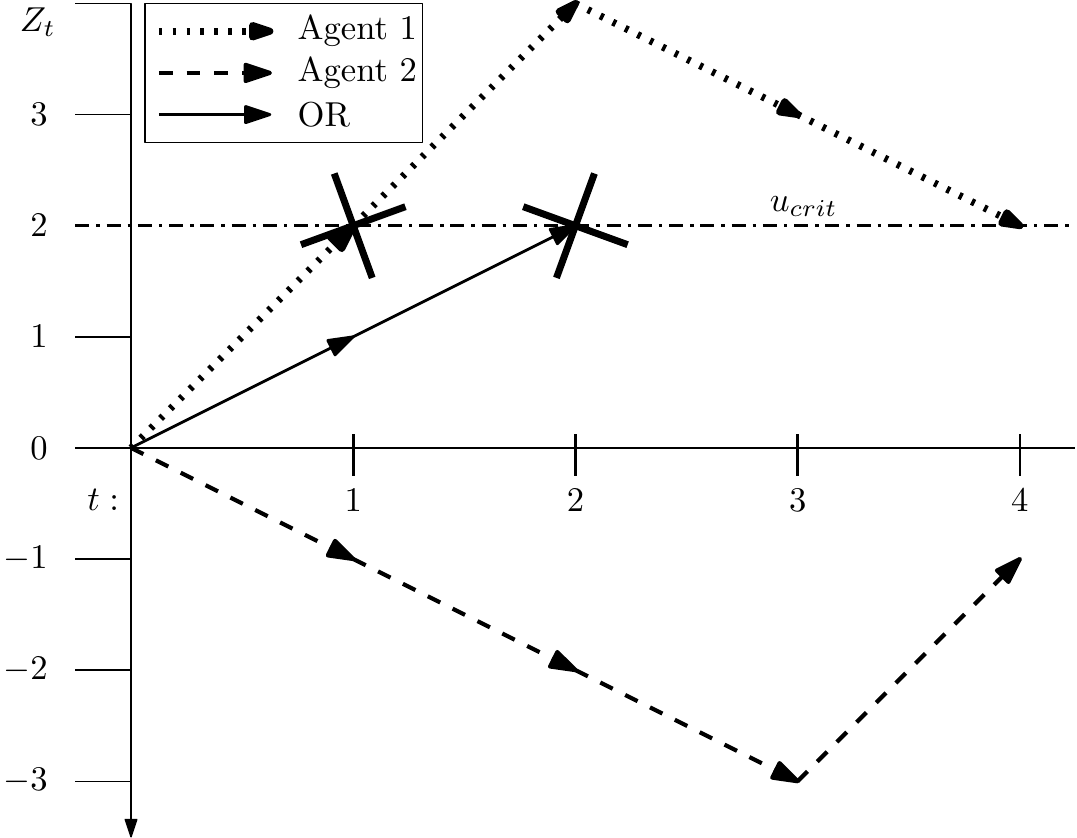}
  \caption{The random walk interpretation of the OR model as well the individual sample paths of the respective agents.}
  \label{fig:DualRW}
\end{figure}

We consider both dual agent models with parameters $c=2$, $r=1$ (and therefore $\theta_\text{crit}=2/3$) and $u_\text{crit}=2$ ($\alpha = 5, \beta=2$). Without determining whether or not these interaction outcomes are witnessed or not, let us presume the responses by the institution to the two agents are given for the first four rounds $(X_{i,1},X_{i,2}, X_{i,3},X_{i,4}) = (0,0,1,1)$ for $i=1$, and $(X_{i,1},X_{i,2}, X_{i,3},X_{i,4}) =(1,1,1,0)$ for $i =2$. The resulting random walk interpretations within the single agent model as well as the observable rewards model are depicted in Figure~\ref{fig:DualRW}.

We observe that the first agent would have quit in the single agent model after the first interaction, but in the OR model they continue past this point. The OA model is not so easily represented visually and so will require more elucidation. We offer Table~\ref{tab:DualRW} summarising the variables over time for the single agent models for agent 1 and 2 respectively as well as in the dual agent models. Note that both agents 1 and 2 have the same estimate throughout in the OR model, due to the fact that they start with the same prior and are privy to the same information each round. The first agent's estimate stops changing after the first round, as a result of their quitting which is the case in the single agent model as well as in the OA dual agent model. This is because by the time they take the decision to not place trust, they have not yet received a signal containing information from agent 2. The reactions of the institution $X_{i,t}$ for $i=1$ and $t\geq2$ are thus not witnessed and not used to update agent 1's belief. 

\begin{table}[htb]
  \centering
  \caption{Estimates, $\hat{\vartheta}_t$ in the respective models.}
  \label{}
  \begin{tabular}{c|c|c|c|c|c|c}
  $t$ & Agent 1 & Agent 2 & OR (1) & OR (2) & OA (1) &OA (2)\\
  \hline
  0 & 0.714 & 0.714 & 0.714 & 0.714 & 0.714  & 0.714 \\
  1 & 0.625 & 0.750 & 0.667 & 0.667 &  0.625 &  0.750\\
  2 &       & 0.778 & 0.636 & 0.636 &        & 0.700\\
  3 &       & 0.800 &       &       &        & 0.727\\
  4 &       & 0.727 &       &      &        & 0.667\\
  \end{tabular}
  \label{tab:DualRW}
\end{table}

In the first round of the OA model both agents behave exactly as they would in the single agent model because their first action is non-informative. After this first round, the first agent quits and takes the ``not place trust'' action in round 2 and all subsequent rounds. This indicates to the second agent that trust was abused in agent 1's first round.
In this case $w_x(1)=1$ (as part of the initial condition of $w_x$) and $I_{2,2}= \theta^0(1-\theta)^1$ while their personal $S_2$ equals $2$, which provides
\begin{equation}
  P_{2,2}(\theta) = \frac{\theta^2 \cdot \theta^{5-1} (1-\theta)^{2-1} \cdot (1-\theta)^1}{\int_0^1y^2 \cdot y^{5-1} (1-y)^{2-1} \cdot (1-y)^1\dY}.
\end{equation}
The mean of this belief distribution is used as the point estimate; it is
\begin{equation}
  \hat{\vartheta}_2 = \int_0^1 \theta\frac{\theta^6 (1-\theta)^{2}}{\int_0^1y^6 (1-y)^{2}\,\dY}\text{d}\theta = 0.7.
\end{equation}
For illustrative purposes we also show the information sent out by agent 2 by their action in round 3, {\it i.e.}, $I_{1,3}$. Firstly they construct $R_2^2$, the belief formed by taking into account the report received by agent 2 by round 2:
\begin{equation}
  R_2^2 (\theta)= \frac{P_0(\theta)I_{2,2}(\theta)}{\int_0^1P_0(y)I_{2,2}(y)\,\dY} =
    \frac{\theta^4(1-\theta)^2}{\int_0^1 y^4(1-y)^2\,\dY}.
\end{equation}
Subsequently the auxiliary distribution $\mathcal{D}_2$ is constructed:
\begin{equation}\label{ex:auxiliary}
  \mathcal{D}_2(x) = \frac{\theta^x (1-\theta)^{2-x}R_2^2(\theta)}{\int_0^1 y^x(1-y)^{2-x}R_2^2(y)\,\dY} = \frac{\theta^{x+4} (1-\theta)^{4-x}}{\int_0^1 y^{x+4}(1-y)^{4-x}\,\dY}.
\end{equation}
This distribution is then used to construct the set $\mathcal{J}_{t=2}^{i=2}=\{2\}$ considering that this is the only $x$ for which $\int_0^1\mathcal{D}_2(x)\,\text{d}\theta \geq 2/3$. Considering that $x=t$ we know that $w_2(2) = 1$, which means that $I_{1,3} = \theta^2$.

This example illustrates an information cascade, as the second agent would not quit in the first 4 rounds in any model except for the OR model. The effect for the first agent on the other hand is that instead of quitting in round 1, as in the single agent model as well as in the OA model, they quit in round 2 of the OR model.

\section{Setup of simulation experiments}\label{sec:Experi}
In this section we describe the setup of our simulations experiments. The results of these simulation experiments we present in \S\ref{sec:res}, while in \S\ref{sec:disc} we discuss and interpret the results of the experiments.
The primary goal of the experiments is to assess the two communication mechanisms in the two agents case, in terms of the probabilities of quitting and the expected quitting times (conditional on quitting). 

\subsection{Choice of \texorpdfstring{$c$}{c} and \texorpdfstring{$r$}{r}}

In order to get an idea of how the model parameters influence the probability (and timing) of agents quitting, we choose the following 5 combinations of $(c,r)$: ${(1,1),(3,2),(2,3),(2,1),(1,2)}$, so that $c=r$ is the ``base case'' and $c=2r$ and $r=2c$ are the ``extremal ends''. There are two reasons for choosing this range of ratios. The first consideration is that if the ratio gets pushed further in either direction, then the simulation results become harder to obtain. The critical trustworthiness $\theta_{\text{crit}}$ is shifted toward $1$ with an increase of $c$, which means that only a small range of $\vartheta$ have probabilities of quitting less than unity. Within this small range there is a sharp drop of quitting probabilities, because with very large $\vartheta$ the probability of quitting is low as this requires an abuse of trust. In the other extreme, with an increase of $r$, $\theta_\text{crit}$ decreases, so that increasingly many parameter settings have to run over a relatively long time interval in an already slow simulation. This is because of the large number of numerical integrations required in the observable actions model used in constructing $\mathcal{J}_t^i$, $\mathcal{D}_t$ and $R_t^i$ for each $i=1,2.$ The second reason is that it is a sufficiently extensive range of cost-to-reward ratios, especially in the context of trust problems\footnote{Equivalent dynamics arise when framing the interaction as paying utility $x$ to the institution. Thereafter an honourable action from the institution equates to receiving utility $x+\delta$. An abusive action on the other hand means the $x$ is lost. Now $c=x$ and $r=\delta$, and our range covers values from $\delta=2x$ to $\delta=x/2$ which we consider trust problems. We consider $\delta>2x$ closer to an issue of gambling, and $\delta<x/2$ closer to transformation of utility from one form to another (without any risk, that is).}.

\subsection{Choice of \texorpdfstring{$\alpha$ and $\beta$}{alpha and beta}}
Intuitively, the parameters $\alpha$ and $\beta$ can be considered as a number of interactions with the institution prior to the dynamics we consider, which resulted in trust being honoured $\alpha$ times and abused $\beta$ times. This means that the greater $\alpha$ is compared to $\beta$, the more the prior belief distribution of the agents is skewed toward greater values of $\vartheta$. Similarly, for a lower $\alpha$ compared to $\beta$, this prior belief distribution is skewed toward the lower values of $\vartheta$. 
The prior belief parameter settings $\alpha$ and $\beta$, in combination with the choice of $c$ and $r$, determine the instance's $u_\text{crit} = r\alpha-c\beta+1$. Borrowing from the analogy of the one-dimensional random walk of the single agent model, the value of $u_\text{crit}$ represents a ``distance'' to quitting. If we would like consistency in the interpretation of $u_\text{crit}$, we require a conversion: Take $u^* = u_\text{crit}/c$, which is the minimum time in which an agent can quit in the single agent model. There is asymmetry in the model between the cases when $c>r$ and $c<r$. In case $c>r$, it is possible for an agent to experience an honouring of trust in the first round, yet still quit after the second if the step size for an abuse of trust is such that $c>u_\text{crit}+r$. In case $r>c$ this scenario is not possible. We choose values of $\alpha$ and $\beta$ such that, in combination with the values of $r$ and $c$, we cover the range $u^* = 1,2,3$. The choice to stop at $u^*=3$ is made to keep the probability of quitting high enough to facilitate efficient simulation.

\subsection{Iterations and simulation length}
In our simulation study, it is our goal to produce reliable estimates for the parameters under study. 
We run each of the models for $4\,000$ truster agents in total, {\it i.e.}, the single agent models are run $4\,000$ times while the dual agent models are run $2\,000$ times. We run each simulation for a maximum of 500 time steps, with the exception of simulations with $\vartheta\in\{0.84,0.9\}$ in which case we run the model for a maximum time of 200 time steps. These choices were made keeping considerations of confidence interval width in mind as well as execution feasibility. Considering that in the two agent models the dynamics are sped up (as there is more information available per time step), this should be sufficient time also for both two agent models.

In the interactions with highest $\vartheta$ the probability of quitting is very low and concentrated on the first couple of time steps. We can make an educated (pessimistic) estimate as to how many time steps would be required in order to not miss any relevant ``probability mass.'' We do this relying on Markov's inequality applied to the expected time to quitting in the single agent model. As an illustration consider the parameter setting with $c=2,r=1, \alpha = 5,$ and $\beta=2$ (such that $u^*=1$) at the parameter setting $\vartheta=0.84$. In the single agent model the expected time to quit is $\mathbb{E}[\tau \,|\,\tau<\infty]\approx 3.17$. By Markov's inequality we know that the probability of quitting at time $\tau\geq 200$ is bounded as follows:
\begin{equation}
  \mathbb{P}(\tau\geq200 \,|\,\tau<\infty) \leq\frac{\mathbb{E}[\tau \,|\,\tau<\infty]}{200} = 0.01587.
\end{equation}
Hence, in the $N$ experiments performed, the expected number of times we miss the quitting event is $L=N\cdot\mathbb{P}(\tau\geq200 \,|\,\tau<\infty)\cdot\mathbb{P}(\tau<\infty)$. Taking $N=4\,000$, we obtain the bound
\begin{equation}
 L < 4000 \times 0.01587 \times 0.2635 = 16.727.
\end{equation}
The single learner case is a conservative benchmark, as there is less communication than in the two agents models, so that we anticipate agents in the two agent models to quit sooner. The extent to which this estimate is conservative, is illustrated by the fact that in the single agent model (for the parameter settings described) the latest quitting occurred at round 45, for the dual agent OA model at round 40, and for the dual agent OR model at round 17.

\section{Results of experiments}\label{sec:res}
\begin{figure}[htb]
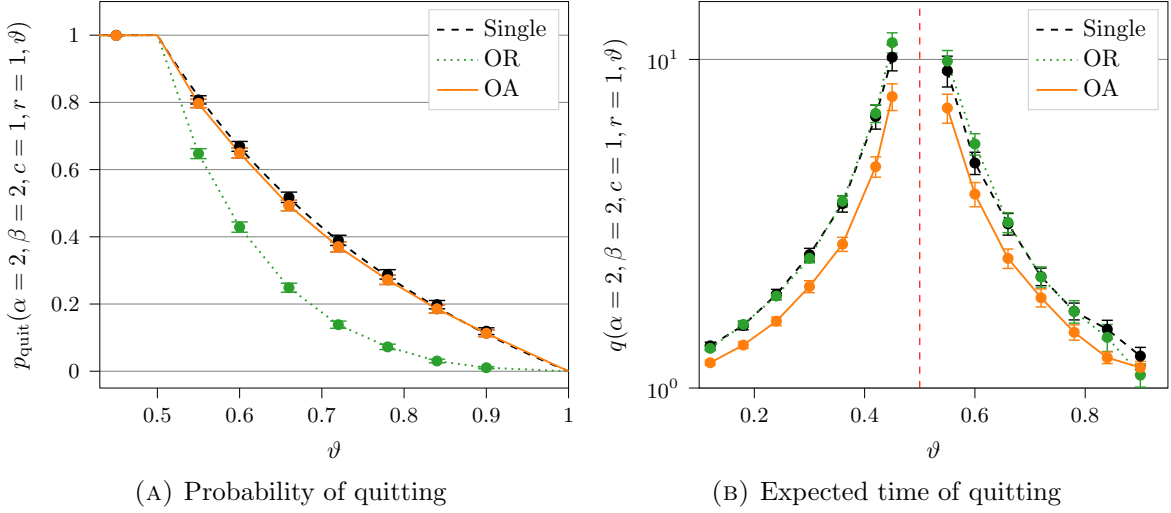

\centering
\subfloat[\centering Probability of quitting]
{ \input{Pi_fig}\label{fig:Qp1q1u1}}
\subfloat[\centering Expected time of quitting]
{\input{Tau_fig}\label{fig:Tp1q1u1}}
\caption{Simulation results for $c=r=1$ in the $u^*=1$ case.}
\end{figure}

In this section we present the results of the experiments described in \S\ref{sec:Experi}. 
We plot the results for {\it specific} values of $(c,r)$ and $u^*$ in Figures~\ref{fig:Qp1q1u1}--\ref{fig:Tp2q1u3}; the numerical output of {\it all} experiments is given, 
in a tabulated format, in Appendix~\ref{AX:tables}. The plots cover the settings
$(c,r)={(1,1),(2,1),(1,2)}$ and $u^*=1,\,3$, where we add the synthetic points $p_\text{quit}(\alpha,\beta,c,r,\vartheta=\theta_\text{crit})=1$ and $p_\text{quit}(\alpha,\beta,c,r,\vartheta=1)=0$ to the curves shown in the figures in order to show the dynamics beyond the capability of the simulation. To be able to see subtle differences in results, we show the probability of quitting predominantly in the region in which quitting is not guaranteed. 

\begin{figure}[htb]
\centering
\subfloat[\centering Probability of quitting]
{ \input{Pi_c1r1u3}\label{fig:Qp1q1u3}}
\subfloat[\centering Expected time of quitting]
{\input{Tau_c1r1u3.tex}\label{fig:Tp1q1u3}}
\caption{Simulation results for $c=r=1$ in the $u^*=3$ case.}
\end{figure}

\begin{figure}[htb]
\centering
\subfloat[\centering Probability of quitting]
{ \input{Pi_c1r2u1.tex}\label{fig:Qp1q2u1}}
\subfloat[\centering Expected time of quitting]
{\input{Tau_c1r2u1.tex}\label{fig:Tp1q2u1}}
\caption{Simulation results for $c=1$ and $r=2$ in the $u^*=1$ case.}
\end{figure}

\begin{figure}[htb]
\centering
\subfloat[\centering Probability of quitting]
{ \input{Pi_c1r2u3.tex}\label{fig:Qp1q2u3}}
\subfloat[\centering Expected time of quitting]
{\input{Tau_c1r2u3.tex}\label{fig:Tp1q2u3}}
\caption{Simulation results for $c=1$ and $r=2$ in the $u^*=3$ case.}
\end{figure}

\begin{figure}[htb]
\centering
\subfloat[\centering Probability of quitting]
{ \input{Pi_c2r1u1.tex}\label{fig:Qp2q1u1}}
\subfloat[\centering Expected time of quitting]
{\input{Tau_c2r1u1.tex}\label{fig:Tp2q1u1}}
\caption{Simulation results for $c=2$ and $r=1$ in the $u^*=1$ case.}
\end{figure}

\begin{figure}[htb]
\centering
\subfloat[\centering Probability of quitting]
{ \input{Pi_c2r1u3.tex}\label{fig:Qp2q1u3}}
\subfloat[\centering Expected time of quitting]
{\input{Tau_c2r1u3.tex}\label{fig:Tp2q1u3}}
\caption{Simulation results for $c=2$ and $r=1$ in the $u^*=3$ case.}
\end{figure}

\subsection{The probability of quitting}
The probability of quitting in the regime $\vartheta<\theta_\text{crit}$ should remain unity: the same and more information is made available to the agents in the dual agent model as in the single agent model. The values of $0.999$ in Tables~\ref{tab:Qu1},~\ref{tab:u2p}, and~\ref{tab:u3p} could have been remedied at the cost of more simulation effort ({\it i.e.}, by working with more runs with more time steps).

In the (more interesting) regime $\vartheta>\theta_\text{crit}$, for most parameter settings the probability of quitting increases from the OR model to the OA model and again to the single agent model. This trend is more pronounced for $u^*=1$ than for $u^*=2$ and $u^*=3$. There are two exceptions to this trend in Tables~\ref{tab:u2p} and~\ref{tab:u3p}: for $\vartheta = 0.65, 0.66$ at the parameter settings $c=3$ and $r=2$, with $u^*=2,3$. However, we observed overlap in the original confidence intervals.

\begin{figure}
\centering
\begin{tikzpicture}[scale = 0.9]

\definecolor{darkgray176}{RGB}{176,176,176}
\definecolor{darkorange25512714}{RGB}{255,127,14}
\definecolor{forestgreen4416044}{RGB}{44,160,44}
\definecolor{gray}{RGB}{128,128,128}
\definecolor{lightgray204}{RGB}{204,204,204}

\begin{axis}[
legend cell align={left},
legend style={fill opacity=1, draw opacity=1, text opacity=1, draw=lightgray204},
tick align=outside,
tick pos=left,
x grid style={darkgray176},
xlabel={\(\displaystyle \vartheta\)},
xmin=0.635, xmax=0.69,
xtick style={color=black},
y grid style={gray},
ylabel={\(\displaystyle q(\alpha=7,\beta=3,c=3,r=2,\vartheta)\)},
ymajorgrids,
ymin=0.45, ymax=0.6,
ytick style={color=black}
]
\addplot [very thick, black, dashed, forget plot]
table {%
0.65 0.556465434225124
0.65 0.587134565774876
};
\addplot [very thick, black, dashed, forget plot]
table {%
0.64 0.556465434225124
0.66 0.556465434225124
};
\addplot [very thick, black, dashed, forget plot]
table {%
0.64 0.587134565774876
0.66 0.587134565774876
};
\addplot [very thick, black, mark=*, mark size=3, mark options={solid}, only marks, forget plot]
table {%
0.65 0.5718
};
\addplot [very thick, black, dashed, forget plot]
table {%
0.66 0.492506822972676
0.66 0.523493177027324
};
\addplot [very thick, black, dashed, forget plot]
table {%
0.65 0.492506822972676
0.67 0.492506822972676
};
\addplot [very thick, black, dashed, forget plot]
table {%
0.65 0.523493177027324
0.67 0.523493177027324
};
\addplot [very thick, black, mark=*, mark size=3, mark options={solid}, only marks, forget plot]
table {%
0.66 0.508
};
\addplot [very thick, forestgreen4416044, dotted, forget plot]
table {%
0.65 0.539055600715307
0.65 0.542144399284693
};
\addplot [very thick, forestgreen4416044, dotted, forget plot]
table {%
0.64 0.539055600715307
0.66 0.539055600715307
};
\addplot [very thick, forestgreen4416044, dotted, forget plot]
table {%
0.64 0.542144399284693
0.66 0.542144399284693
};
\addplot [very thick, forestgreen4416044, mark=*, mark size=3, mark options={solid}, only marks, forget plot]
table {%
0.65 0.5406
};
\addplot [very thick, forestgreen4416044, dotted, forget plot]
table {%
0.66 0.475852067617446
0.66 0.478947932382554
};
\addplot [very thick, forestgreen4416044, dotted, forget plot]
table {%
0.65 0.475852067617446
0.67 0.475852067617446
};
\addplot [very thick, forestgreen4416044, dotted, forget plot]
table {%
0.65 0.478947932382554
0.67 0.478947932382554
};
\addplot [very thick, forestgreen4416044, mark=*, mark size=3, mark options={solid}, only marks, forget plot]
table {%
0.66 0.4774
};
\addplot [very thick, darkorange25512714, forget plot]
table {%
0.65 0.526041212554232
0.65 0.532958787445768
};
\addplot [very thick, darkorange25512714, forget plot]
table {%
0.64 0.526041212554232
0.66 0.526041212554232
};
\addplot [very thick, darkorange25512714, forget plot]
table {%
0.64 0.532958787445768
0.66 0.532958787445768
};
\addplot [very thick, darkorange25512714, mark=*, mark size=3, mark options={solid}, only marks, forget plot]
table {%
0.65 0.5295
};
\addplot [very thick, darkorange25512714, forget plot]
table {%
0.66 0.466241544662975
0.66 0.473158455337025
};
\addplot [very thick, darkorange25512714, forget plot]
table {%
0.65 0.466241544662975
0.67 0.466241544662975
};
\addplot [very thick, darkorange25512714, forget plot]
table {%
0.65 0.473158455337025
0.67 0.473158455337025
};
\addplot [very thick, darkorange25512714, mark=*, mark size=3, mark options={solid}, only marks, forget plot]
table {%
0.66 0.4697
};
\addplot [very thick, black, dashed]
table {%
0.1 0
0.2 0
};
\addlegendentry{Single}
\addplot [very thick, darkorange25512714]
table {%
0.1 0.1
0.2 0.1
};
\addlegendentry{OA}
\addplot [very thick, forestgreen4416044, dotted]
table {%
0.1 0.2
0.2 0.2
};
\addlegendentry{OR}
\end{axis}

\end{tikzpicture}
\caption{The results of extra simulation runs for the probability of quitting. In these runs $c=3$, $r=2$, and $u^*=2$.}\label{fig:Extra}
\end{figure}
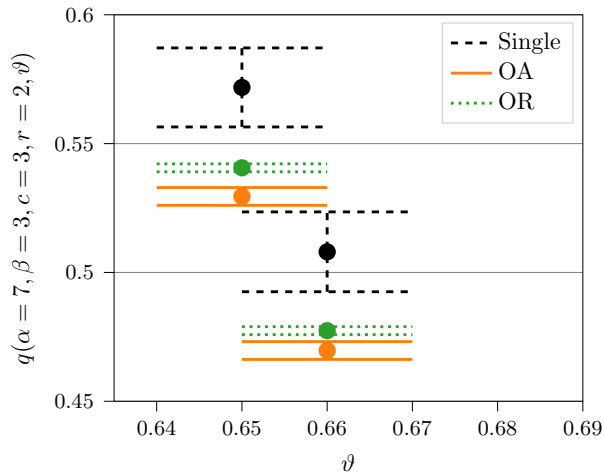

To determine whether such a difference bears significance we conducted more simulations for the faster case with $u^*=2$. For the OR model, being computationally lighter, we conducted a total of 200\,000 runs. For the slower OA model, we conducted 40\,000 simulations in total. The resulting confidence intervals of the points $\vartheta=0.65, 0.66$ are depicted in Figure~\ref{fig:Extra}. We observe the, perhaps unexpected, result that the OA model produces a {\it lower} probability of quitting in this parameter setting. Hence, the OR model does not \textit{always} outperform the OA model in terms of the probability of quitting in the regime where $\vartheta>\theta_\text{crit}$.

Part of the explanation of the lower probability of quitting in the OA model relates to the timing of the communication, being {\it at the end} of a round. In the OR model, as soon as an agent has their own reward, they also observe the reward of their neighbour. This boils down to two pieces of information per round. In the OA model however, the agents only observe the actions of their neighbour at the same time when they have already made their own action for that round. That means that an agent who observes their neighbour not placing trust for the first time in round $t$ can only use the information contained therein to inform their action in round $t+1$. In the meantime, as they have not quit yet in round $t$, they are privy to at least that round's outcome before they have to make another decision. We elaborate on this effect in Appendix~\ref{AX:illustration}.

\subsection{Expected time to quit}
In general (in the regime $\vartheta<\theta_{\text{crit}}$) the expected time to quit is lower in the dual agent models than in the single agent model. Furthermore, the OR model tends to have the lowest expected time to quit ({\it i.e.}, the greatest effect), though with a fair amount of crossing with the OA model. The key exception to this trend occurs at $c=r=1$ with $u^*=1$, depicted in Figure~\ref{fig:Tp1q1u1}. At this setting, the OA model performs better than both others which show a similar time to quitting. Furthermore, there are exceptions at $c=2$, $r=1$ and $u^* = 1$ depicted in Figure~\ref{fig:Tp2q1u1}, and at $c=3$, $r=2$ and $u^* = 1$ shown in Table~\ref{tab:Tu1}. 

In the regime $\vartheta<\theta_\text{crit}$ there is the trend (with exceptions) that quitting in the OR model occurs sooner than in both other models. In the OR model the agents are exposed to two outcomes per round of interaction and so receive twice as much unbiased information than in the single agent model. In this regime, the unbiased information thus received is likely to indicate that the institution is trustworthy and so quitting becomes less likely with more time ({\it i.e.}, quitting must occur quickly or not at all). This trend is most pronounced when $u^*=3$ shown in Figures~\ref{fig:Tp1q2u3} and~\ref{fig:Tp2q1u3} as well as in Table~\ref{tab:u3t}.

\section{Discussion}\label{sec:disc}
In this section we draw conclusions from the numerical experiments, discuss how these findings relate to those in the literature, and present directions for future research.
\subsection{General conclusions}
In all of the models a similar pattern in the expected time to quitting holds: there is an increase in the expected time to quit as $\vartheta$ increases until $\theta_{\text{crit}}$ and a decrease afterwards. This specifically entails that a long average tenure of customers at an institution, does not indicate that this institution is to be trusted, or, put differently, there is no way of knowing which side of the critical value they might be on with only this information. An institution can simply have a trustworthiness that is just high enough to keep customers placing their trust in them long enough to get positive net utility from that relationship yet not actually high enough to warrant an indefinite relationship. An institution with many relationships still going and a couple of very short concluded ones on the other hand might be a good indication that the institution is indeed trustworthy.

By comparing the probability of quitting for the same $c$ and $r$ at different $u^*$ (for example in Figures~\ref{fig:Qp1q1u1} and~\ref{fig:Qp1q1u3}) we see that the difference between the three models is relatively small for $u^*=2,3$. This shows that starting optimistically allows agents to secure good chances of trusting a trustworthy institution in the long run. 

We encountered the (perhaps) remarkable phenomenon that the effects of the different two agent communication mechanisms are not monotone. This is highlighted in the cases (a) with $c=r=1$ and $u^*=1$, depicted in Figures~\ref{fig:Qp1q1u1} and~\ref{fig:Tp1q1u1} and (b), the case with $c=3$, $r=2$ and $u^*=2$, depicted in Figure~\ref{fig:Extra}. In (a), the OR model outperforms the OA model in terms of probability of trusting a trustworthy institution. At the same time however, the OA model outperforms the OR model in making the decision to quit a trust relationship with an untrustworthy institution sooner. In (b), we see the OA model outperforming the OR model in terms of the probability of trusting a trustworthy institution for values $\vartheta=0.65,0.66$. These results show us that we cannot state that the OR model is always ``better'' than the OA model. In fact this depends on the parameter setting as well as on which criterion you find most important: making the correct decision in the long run, or being sure to end a relationship with an untrustworthy institution relatively quickly.

This surprising result is partially due to the ``timing'' of the underlying dynamics. Agents make their decision to place trust at the same time and so can only use information from their neighbours action, one round later. We illustrate this in a two-rounds setting in Appendix~\ref{AX:illustration}.

We summarise our insights from the numerical results:
\begin{itemize}
    \item Communication always helps: it increases the \emph{probability} of never ceasing to trust a trustworthy institution, and it decreases the \emph{expected time until quitting} a relationship with an untrustworthy institution, both compared to the single agent model. 
    
   \item The OA model can be better or worse than the OR model, depending on the performance measure of interest (probability of quitting, or expected time to quit). There are instances in which having less information is beneficial to the agents.  
    \item A good way to increase chances of trusting a trustworthy institution in the long run without a social network is to start with a more optimistic prior. 
\end{itemize}

\subsection{Reflection and context} 
In this paragraph we compare our model to those found in the literature streams conceptually. In the subsequent paragraphs we compare the outcomes of our model to those in the literature. We presented a model of trust which includes both learning from interactions with the institution as well as learning from communication between agents. Thematically this work is related to social network trust where the focus is on the possible loss of trust. Methodologically this work is related to the social network learning. The agents in our model are learning about the environment (trustworthiness of the trustee). Our model extends ideas in both of these streams in a natural way: The dependence structure between actions taken by the agents, and the signals they subsequently use to learn, extend in a realistic way the work from the social network learning literature, which typically does not cover this complication. Finally, the observable actions model of communication adds realism to the relationships between agents, not present in the more common observable rewards model of communication in the social network trust literature. 

The results of our investigation, akin to those from Buskens~\cite{Buskens2003}, show that trust increases in the models with truster agents sharing information about the trustee. In our model, however, this effect extends to models in which a trustee dishonouring trust in one round does not immediately indicate that they are not to be trusted. Furthermore, we extended the type of communication between the truster agents beyond complete information sharing. We find that the positive effect of agent communication, though stronger in the complete information sharing, is also present in the model in which communication is limited to observing actions. Hence, even when individuals do not hold extensive discussion with their peers about the institution, simply observing their actions provides a significant benefit, which has previously been shown for more extensive communication in the social network trust literature~\cite{Buskens2003,Frey2015,Buskens2010}.

In both models of communication that we consider, for all parameter settings, we observe that rational use of social network information increases the chances of agents to learn the true trustworthiness when it is rational to place trust. We also observe that for both models of communication and in all parameter settings, the expected time to quitting is sooner, which is especially beneficial when it is rational to quit. Our model includes realistic assumptions in terms of agent communication and of signal dependence on actions, and shows that communication between agents are an aid to learning and trusting. The simplistic nature of independent signals and actions cannot be imported to models of trust and learning. It is the nature of a trust problem that resources have to be placed at the liberty of the trustee agent in order to see how they respond. However, like in the social network learning literature (cf.~\cite{Harel2021,Huang2022}), our work shows that more communication leads to \textit{faster} dynamics, but, due to the dependence of signals on actions, sometimes the result is converging to the sub-optimal action (i.e., not placing trust).

\subsection{Future work}
We see from the developed model that agents communicating with one another bears a benefit. There are, however, situations in which the agent that otherwise might have stayed on the course of placing trust, ``wrongfully'' stops placing trust as a result of what they hear from their neighbour. It seems that the overall dynamics are not dominated by this effect. A natural question then becomes whether or not this beneficial effect retains in a model with more than two agents. The present model is limited to two agents partially as a result of the intense computational work involved: agents in the OA model perform numerous numerical integrations per round in order to identify which histories of experiences are plausible for the communication received from their neighbour. It is an open question whether there is a scalable approach to perform these computations. Alternatively, one could find a way around this by relaxing the agents' rationality when it comes to interpreting their neighbours actions. This would allow investigation of a greater pool of agents who have some sophistication in learning from their private signal, but comes at the cost of simplifying the model. 

Another line of future work concerns asymmetric communication between two truster agents. One of the truster agents may be modelled as a news outlet which only sends information without receiving any in return. In the same spirit one agent might be malicious by spreading misinformation about the institution. It may be that this asymmetry between agents may make technical results more attainable. For instance, one can condition on the private signal of the only sending truster and observe the dynamics of the receiving truster. The dynamics for the purely sending truster agent conveniently follows the dynamics we have presented in the context of the single-agent trust model. Modelling a malicious actor is also possible by deciding a priori what signal they will send, or by having a truthfulness parameter by which the truster agent communicates their actual experience with probability $\eta$ and communicates that trust was abused (regardless of the truth) with probability $1-\eta$. The honest agent in such a model may then need to learn not only about the trustworthiness of the institution but also about the information they receive from their network.

\bibliographystyle{ieeetr}
\bibliography{Biblio.bib}

\setcounter{section}{0}
\renewcommand\thesection{\Alph{section}}
\section{Appendix: Proofs of results}\label{AX:proofs}

\subsection{Proof of Lemma~\ref{lem:quit}(Guaranteed quitting)}
\begin{proof}
We prove this by contradiction. Suppose that $A_t=1$ for all $t\in\mathbb{N}$. Hence,
\begin{equation}
  r\hat{\vartheta}_t-c(1-\hat{\vartheta}_t)\geq 0, \quad \forall t\in\mathbb{N}.
\end{equation}
The agent estimates the trustworthiness in round $t$ by
\begin{equation}
  \hat{\vartheta}_t=\frac{\alpha+\hat{S}_t}{\alpha+\beta+t},
\end{equation}
and we know by the law of large numbers that 
\begin{equation}
  \lim_{t\to \infty}\frac{S_t}{t} = \vartheta \quad \text{(almost surely).}
\end{equation}
This also implies that 
\begin{equation}
  \lim_{t\to\infty}\hat{\vartheta}_t = \vartheta <c/(r+c) \quad \text{(almost surely).}
\end{equation}
This means that there exists $t$ for which
\begin{align}
  r\hat{\vartheta}_t-c(1-\hat{\vartheta}_t)&<r\left(\frac{c}{c+r}\right)-c\left(1-\frac{c}{c+r}\right),\\
  &<\frac{rc}{c+r} - \frac{c(c+r)}{c+r}+\frac{c^2}{c+r}= \frac{0}{c+r}.
\end{align}
At this value of $t$, the expected utility of the interaction is negative, which implies that the agent takes action $A_t = 0$, a contradiction.
\end{proof}

\subsection{Proof of Lemma~\ref{RorC} (Converge or quit)}
\begin{proof}
The agent makes use of a Bayesian belief update starting with a beta distributed prior belief with shape parameters $\alpha$ and $\beta$. Given that $\hat{S}_t=\sum_{t=1}^{t}X_t\mathbf{1}_{\{A_t=1\}}$, where $X_t$ is a binary random variable taking the value 1 at probability $\vartheta$ and $0$ at probability $(1-\vartheta)$, where $\vartheta$ is the value which the agent is trying to estimate, we have that the belief distribution at time $t$ equals
\begin{equation}
  P_t(\theta) = \frac{\theta^{\hat{S}_t+\alpha-1}(1-\theta)^{(t-\hat{S}_t)+\beta-1}}{\int_0^1y^{\hat{S}_t+\alpha-1}(1-y)^{(t-\hat{S}_t)+\beta-1}\dY},\quad \theta \in [0,1].
\end{equation}
Defining $\tau$ as the first $t\in\mathbb{N}$ such that $\int_0^1\theta P_t(\theta)\,\dT<c/(c+r)$ (possibly $\infty$), we have for all $t<\tau$ that the estimated $\vartheta$ is given by
\begin{equation}
  \hat{\vartheta}_t = \frac{\alpha+\hat{S}_t}{\alpha+\beta+t}.
\end{equation}
Then there are two possible cases: $\tau<\infty$ and $\tau = \infty$. In the first case (where $\tau<\infty$) the agent stops placing trust and by the assumption on $\vartheta$, namely that $\vartheta\geq c/(c+r)=\theta_\text{crit}$, it is clear that the agent has not converged to the true value of $\vartheta$. This is because of the fact that quitting requires $\hat{\vartheta}\geq\theta_\text{crit}$ and that by assumption we have chosen $\vartheta>\theta_\text{crit}$. In the second case we have $\tau = \infty$. Here we define $S_t$ which drops the dependence on the actions of the agent:
\begin{equation}
  S_t=\sum_{s=1}^tX_t,\quad \forall t\in\mathbb{N},
\end{equation} 
and we note that when $\tau = \infty$ the relationship $\hat{S}_t=S_t$ holds, and therewith $\hat{\vartheta} = (S_t+\alpha)/(\alpha + \beta + t)$. First inspecting $S_t/t$ which we know by the law of large numbers takes
\begin{equation}
  \lim_{t\to\infty}\frac{S_t}{t} = \vartheta\quad\text{(almost surely)}.
\end{equation}
Subsequently investigating $\hat{\vartheta}_t$ as $t\to\infty$ we see:
\begin{equation}
 \lim_{t\to\infty} \hat{\vartheta}_t=\lim_{t\to\infty}\frac{S_t+\alpha}{t+\alpha+\beta} = \vartheta \quad \text{(almost surely)}.
\end{equation}
This shows that the agent's estimate has indeed converged to the true $\vartheta$ almost surely.
\end{proof}

\subsection{Proof of Lemma~\ref{absp1} (Quitting probability when \texorpdfstring{$c=1$}{c=1}, and \texorpdfstring{$r\in\mathbb{N}$}{r in N})}
\begin{proof}
Recall the assumption that $\theta\geq c/(c+r)$ and the definitions  $\pi(u) = \mathbb{P}(\exists t: Z_t \geq u)$ and $\varrho = \mathbb{P}(\exists t : Z_t \geq 1)$. Observe that (A) by the strong Markov property the random walk after having hit level 1 is independent from what happened before, (B) on the path to level $u$ any level in the set $\{1,\ldots,u-1\}$ has been attained. We thus obtain the `memoryless property'
\begin{equation}\label{piu}
  \pi(u)=\varrho^u.
\end{equation}
Now consider the first step of the random walk starting at 0; it can either hit the first level immediately at probability $(1-\theta)$ or it drops to the level $-r$ putting it a distance $r+1$ from the level 1. Using (\ref{piu}) we find the identity
\begin{equation}\label{rhoSol}
 F(\varrho):=  \varrho = (1-\vartheta) + \vartheta \varrho^{r+1}=:G(\varrho).
\end{equation}
We can prove that this equation has exactly one solution in the interval $[0,1)$. Note that $F(\cdot)$ is convex. A trivial solution is of course $\varrho = 1$ which accounts for one of at most two intersections between a convex function and a linear function. It thus remains to show that there is another intersection, and that it lies in $[0,1)$. This is indeed the case because $F'(1)=\theta(r+1)$, which is larger than $G'(1)=1$ due to the condition that $\theta> c/(c+r)$ after substituting $c=1$ and rearranging. Observing that $F(0)=1-\theta>0=G(0)$, this entails that there is another root located somewhere between $\varrho=0$ and $\varrho =1$.
\end{proof}

\subsection{Proof of Lemma~\ref{lem:exTqN} (Expected time to quitting when \texorpdfstring{$c=1$}{c=1}, and \texorpdfstring{$r\in\mathbb{N}$}{r in N})}
\begin{proof}
Let $\tau(u)=\inf\{t:Z_t = u\}$ be the first passage time for the walk $Z_t$ to $u$ and recall that $\pi(u)$ is the probability of this event occurring in finite time (\textit{i.e.},\,$\pi(u) :=\mathbb{P}(\exists t :Z_t\geq u)$). Then we are interested in the expectation of the $\tau(u)$ conditioned on its finity:
\begin{equation}
  \mathbb{E}[\tau(u) \,|\,\tau(u)<\infty] = \frac{\mathbb{E}[\tau(u)\mathbf{1}\{\tau(u)<\infty\}]}{\pi(u)}.
\end{equation}
Note the relationship
\begin{equation}\label{taurel}
  \mathbb{E}[z^{\tau(u)}\mathbf{1}\{\tau(u)<\infty\}] = \mathbb{E}[z^{\tau(1)}\mathbf{1}\{\tau(1)<\infty\}]^u
\end{equation}
which holds by the precise same `memoryless argumentation' as in the proof of Lemma~\ref{absp1}. We define 
\begin{equation}
  \varphi(z):= \mathbb{E}[z^{\tau(1)}\mathbf{1}\{\tau(1)<\infty\}],
\end{equation}
which by the relationship (\ref{taurel}) also allows for a self-referential expression 
\begin{equation}
  \varphi(z) = z (1-\vartheta) + z \vartheta\, \varphi(z)^{r+1}.
\end{equation}
Finally note that we can get the expectation of interest out again by taking the derivative and substituting $z=1$:
\begin{equation}
  \mathbb{E}[\tau(u) \,|\,\tau(u)<\infty] = u\frac{\varphi'(1)}{\pi(1)}.
\end{equation}
We now manipulate these expression in order to get the final statement: observing that
\begin{equation}
  \varphi'(z) = (1-\vartheta) + \vartheta\varphi(z)^{r+1} + z\vartheta(r+1)\varphi(z)^{r}\varphi'(z),
\end{equation}
substituting $z=1$ and noting that $\varphi(1)=\pi(1)=\varrho$ with $\varrho$ as defined in (\ref{rhodef}), we obtain
\begin{equation}
\begin{split}
  \varphi'(1) &= \underbrace{(1-\vartheta) +\vartheta\varrho^{r+1}}_{=\varrho \text{ by (\ref{eq:rhosol})}} + \vartheta (r+1) \varrho^{r}\varphi'(1)=\frac{\varrho}{1-(r+1)\underbrace{\vartheta\varrho^r}_{=1-(1-\vartheta )\varrho^{-1}}}\\
  &=\frac{\varrho^2}{(1-\vartheta)(r+1)-r\varrho}\quad (\text{multiplying by } \frac{\varrho}{\varrho}).
  \end{split}
\end{equation}
Multiplying this by $u/\varrho$ gives the final result.\end{proof}

\subsection{Proof of Lemma~\ref{lemq1} (Quitting probability when \texorpdfstring{$r=1$}{r=1}, and \texorpdfstring{$c\in\mathbb{N}$}{c in N})}
\begin{proof}

We introduce the time $G$ which follows a shifted geometric distribution: $\mathbb{P}(G=t)=(1-f)^tf$ for some $f\in(0,1)$. We investigate the behaviour of the walk, specifically the probability of reaching $u$, before the geometrically triggered time $G$:
\begin{equation}
  p[f](u):=\mathbb{P}(\exists n\leq G : Z_n\geq u),
\end{equation}
as well as the probability of reaching $Z_n=1$ for some $n\leq G$:
\begin{equation}
  \varrho[f]:=\mathbb{P}(\exists n \leq G :Z_n\geq 1).
\end{equation}
Furthermore, we define minimum of the walk until time $n$, $\underline{Z}_n:=\min \{Z_0, Z_1,\ldots,Z_n\}$ as well as the maximum attained until time $n$, $\bar{Z}_n:=\max \{Z_0, Z_1,\ldots,Z_n\}$. Due to the Wiener-Hopf factorisation \cite{KYP, KYP2}, $\bar{Z}_G-Z_G$ is independent of $\bar{Z}_G$, so that
$\mathbb{E}w^{Z_G}=\mathbb{E}w^{\bar{Z}_G}\mathbb{E}w^{Z_G-\bar{Z}_G}.$ It is also true that $Z_G-\bar{Z}_G$ is distributed as $\underline{Z}_G$, as is seen by rotating the walk $180^\circ$ and considering $Z_G$ as the starting point of the new, identically distributed process. Therefore,
\begin{equation}\label{eq:WH}
   \mathbb{E}w^{Z_G}=\mathbb{E}w^{\bar{Z}_G}\mathbb{E}w^{\underline{Z}_G}.
\end{equation}
Hence, if we manage to identify $\mathbb{E}w^{Z_G}$ and $\mathbb{E}w^{\underline{Z}_G}$, we have found $\mathbb{E}w^{\bar{Z}_G}$. In pursuit of $\mathbb{E}w^{Z_G}$, let $x$ be the number of times the process jumps upward in the time interval $t\in\{1, \ldots,G\}$. Hence,
\begin{equation}
\begin{split}
  \mathbb{E}w^{Z_G}&=\mathbb{E}\left[\sum_{x=0}^G\binom{G}{x}(1-\vartheta)^x\vartheta^{G-x}w^{cx-(G-x)} \right]\\
  &= \mathbb{E}\left[ \sum_{x=0}^G\binom{G}{x}((1-\vartheta)w^c)^x\left(\frac{\vartheta}{w}\right)^{G-x}\right]=\mathbb{E}\left[(1-\vartheta)w^c+\vartheta w^{-1}\right]^G.
\end{split}
\end{equation}
Recalling that $G$ is geometrically distributed, for $|(1-f)((1-\theta)w^c+\theta w^{-1})|<1$,
\begin{equation}\label{eq:EwZ}
\begin{split}
 \mathbb{E}w^{Z_G}&=\sum_{i=0}^\infty (1-f)^i f((1-\vartheta)w^c+\vartheta w^{-1})^i
  =\frac{f}{1-(1-f)((1-\vartheta)w^c+\vartheta w^{-1})}.
 \end{split}
\end{equation}

In order to find $\mathbb{E}w^{\underline{Z}_G}$ we note that the minimum $\underline{Z}_G$ is identically distributed to the negative of the maximum of the random walk $B_G$ that takes steps up of size one with probability $\vartheta$ and down $c\in \mathbb{N}$ with probability $1-\vartheta$. These two walks and their step descriptions are shown in Table~\ref{tab:2walks}.
\begin{table}[ht]
\begin{tabular}{l|l|l|l|l}
Walk & Step $+$ & ${\mathbb P}(+)$ & Step $-$ & ${\mathbb P}(-)$\\
\hline
$Z_G$ & $c$ & $1-\vartheta$ & 1 & $\vartheta$\\
$B_G$ & 1 & $\vartheta$ & $c$ & $1-\vartheta$
\end{tabular}
\caption{Two random walks.}
\label{tab:2walks}
\end{table}
The walk $B_G$ is simply a variable substitution of the walk analysed in in \S\ref{secp1} at an added shifted geometric distributed time step $G$ and as such the analysis is remarkably similar. Define the probability of the walk $B$ going up by at least $u$ before the geometric trigger,
\begin{equation*}
  p[f](u):=\mathbb{P}(\exists n\leq G: B_n\geq u).
\end{equation*}
Similarly define the probability of reaching $u=1$ before the geometric trigger by
\begin{equation*}
  \varrho[f]:=\mathbb{P}(\exists n\leq G : B_n\geq 1),
\end{equation*}
then as before we have the memoryless property $p[f](u) = (\varrho[f])^u$.
It is also apparent that $\varrho[f]$ solves
\begin{equation} \label{rhoGSolve}
  \varrho[f] = (1-f)(\vartheta +(1-\vartheta)(\varrho[f])^{c+1}).
\end{equation}

By defining $s[f](u):=\mathbb{P}(\bar{B}_G = u)$ as the probability that the maximum of the random walk $B$ by time $G$ is exactly $u$,
and noting that 
\begin{equation}
\begin{split}
  s[f](u) &= p[f](u) - p[f](u+1) = (\varrho[f])^u - (\varrho[f])^{u+1} = (\varrho[f])^u(1-\varrho[f]),
\end{split}
\end{equation}
we have an expression for this maximum. By the relationship between $B_G$ and $Z_G$, we have also have an expression for $\underline{Z}_G$, with $\varrho[f]$ solving (\ref{rhoGSolve}),
\begin{equation*}
  \mathbb{P}(\underline{Z}_G=-u) = (\varrho[f])^u(1-\varrho[f]).
\end{equation*}
Hence, 
\begin{equation}
  \mathbb{E}w^{\underline{Z}_G}= \sum_{u=0}^\infty(\varrho[f])^u(1-\varrho[f])w^{-u}= \frac{1-\varrho[f]}{1-\varrho[f] w^{-1}}.
\end{equation}
This, together with expression (\ref{eq:EwZ}) and the Wiener-Hopf factorisation (\ref{eq:WH}), yields
\begin{equation}
  \mathbb{E}w^{\bar{Z}_G} = \frac{f}{1-(1-f)((1-\vartheta)w^c+\vartheta w^{-1})}\times \frac{1-\varrho[f]w^{-1}}{1-\varrho[f] }.
\end{equation}
By considering this as $f\downarrow 0$, it is first noted that $\varrho[0]=1.$
Then, defining $\sigma(u):= {\mathbb P}(\bar Z_\infty = u)$,
\begin{equation}
\begin{split}
  \zeta(w)&:=\sum_{u=0}^{\infty}\sigma(u)w^u=\lim_{f\downarrow 0}\mathbb{E}w^{\bar{Z}_G} \\
  &=\lim_{f\downarrow 0} \frac{f}{1-(1-f)((1-\vartheta)w^c+\vartheta w^{-1})}\times \frac{1-\varrho[f]w^{-1}}{1-\varrho[f] }\\
  &=\lim_{f\downarrow 0}\frac{1-\varrho[f]w^{-1}}{1-(1-f)((1-\vartheta)w^c+\vartheta w^{-1})}\times \frac{f}{1-\varrho[f]}\\
  &=\frac{1-w^{-1}}{1-((1-\vartheta)w^c+\vartheta w^{-1})}\times\lim_{f\downarrow 0} \frac{f}{1-\varrho[f]}
\end{split}
\end{equation}
Applying L'H\^{o}pital's rule,
we obtain 
\begin{equation}
  \zeta(w) = \frac{1-w^{-1}}{1-((1-\vartheta)w^c+\vartheta w^{-1})}\times\frac{1}{-\gamma},\:\:\:\:\mbox{with}\:\:\:\gamma:= \lim_{f\downarrow 0}\frac{\text{d}}{\text{d}f}\varrho[f].
\end{equation}
It is easy to verify, by differentiating both sides of (\ref{rhoGSolve}) with respect to $f$ and solving for $\gamma$, that $1/\gamma = {(c+1)(1-\vartheta)-1},$ which entails that
\begin{equation}\label{eq:xi_as_sigma}
\begin{split}
  \zeta (w) &= \frac{(1-w^{-1})((c+1)(1-\vartheta)-1)}{((1-\vartheta)w^c+\vartheta w^{-1})-1}=\frac{(w-1)(1-(1-\vartheta)(c+1))}{w-\vartheta-(1-\vartheta)w^{c+1}}.
\end{split}
\end{equation}
But we are interested in $\pi(u)$, as such let $\xi(w)=\sum_{u=0}^{\infty}\pi(u)w^u$, which is expressed using $\sigma(u)$ in the following way
\begin{equation}
\begin{split}
  \xi(w) &= \sum_{u=0}^{\infty}\sum_{v=u}^\infty\sigma(v)w^u= \sum_{v=0}^{\infty}\sum_{u=0}^v\sigma(v)w^u\\
  &= \sum_{v=0}^{\infty}\sigma(v)\left(\frac{1-w^{v+1}}{1-w}\right) = \frac{\sum_{v=0}^\infty\sigma(v) - w\sum_{v=0}^\infty \sigma(v)w^v}{1-w},
\end{split}
\end{equation}
culminating in the useful expression
\begin{equation}
  \xi(w) = \frac{1-w\zeta(w)}{1-w},
\end{equation}
by which we can evaluate $\pi(u)$ via the relationship
\begin{equation}
  \pi(u)=\left.\frac{1}{u}\frac{\text{d}^u}{\text{d}w^u}\xi(w)\right|_{w=0} = \frac{\xi^{(u)}(0)}{u!}.
\end{equation}
Substituting the expression for $\zeta(w)$ gives the result.\end{proof}

\remark{Cong and Sato~\cite{CongSato1982} use combinatorial arguments to find the probability of first hitting the absorbing barrier at specific times $t$ for the cases when either $r$ or $c$ is one,
\begin{equation}
  p_t(u)= \mathbb{P}( \inf\{n\in \mathbb{N}: Z_n\geq u\}=t).
\end{equation}
Taking the sum (over the first hitting times $t$) could in turn provide the probabilities that we state here: $\sum^{\infty}_{t=1}p_t(u)=\pi(u)$. Notice that this approach is of an intrinsically approximative nature, as one has to somehow truncate the summation over $t$, while our approach is exact. We do not make use of their result, also because the underlying computations are involved, and in addition it does not yield the expected time of quitting.}

\subsection{Proof of Lemma~\ref{lem:exTpN} (Expected time to quitting when \texorpdfstring{$r=1$}{r=1}, and \texorpdfstring{$c\in\mathbb{N}$}{c in N})}
\begin{proof}
Define
\begin{equation}
  \varphi(z,u):=\mathbb{E}[z^{\tau(u)}\mathbf{1}\{\tau(u)<\infty\}],\:\:\:
  \Phi(w,z):=\sum_{u=1}^{\infty}w^u\varphi(z,u), \quad w\in(-1,1),
\end{equation}
so that
\begin{equation}
  \mathbb{E}[\tau(u) \,|\,\tau(u)<\infty] = \frac{\varphi'(1, u)}{\varphi(1,u)},
\:\:\:\:
  \varphi(z,u)= \left.\frac{\partial^u}{\partial w^u}\frac{\Phi(w,z)}{u!}\right|_{w=0}
\end{equation}

The dynamics of the process reveal the recursive relation
\begin{equation}
\varphi(z,u) = \begin{cases}
(1-\vartheta)z\,\varphi(z,u-c) + \vartheta z\, \varphi(z,u+1) \quad &\text{if }u>c,\\
 (1-\vartheta)z + \vartheta z \,\varphi(z,u+1) &\text{otherwise.}
\end{cases}
\end{equation}
Substituting this into the definition of $\Phi(w,z)$, we obtain, after some minor rewriting,
\begin{equation}
  \begin{split}
    \Phi(w,z)=&\underbrace{\sum_{u=1}^c w^u (1-\vartheta)z}_{(a)} + \underbrace{\sum_{u=1}^\infty w^u \vartheta z \,\varphi(z,u+1)}_{(b)}+\underbrace{\sum_{u=c+1}^\infty w^u(1-\vartheta)z\,\varphi(z,u-c)}_{(c)}.
  \end{split}
\end{equation}
Inspecting $(a)$, $(b)$, and $(c)$ individually, we obtain
\begin{align}
\begin{split}
  (a) &=(1-\vartheta)z \sum_{u=1}^c w^u=(1-\vartheta)z \left(\frac{w-w^{c+1}}{1-w}\right)
\end{split}
\end{align}
by applying the finite geometric series, 
\begin{align}
\begin{split}
    (b) &= \vartheta z \frac{1}{w}\sum_{u=1}^\infty w^{u+1}\varphi(z,u+1)=\frac{\vartheta z}{w}\Phi(w,z)-\vartheta z \,\varphi(z,1)
\end{split}
\end{align}
relying on the definition of $\Phi(w,z)$, and
\begin{align}
\begin{split}
    (c)&=w^{c}(1-\vartheta) z \sum_{u=c+1}^\infty w^{u-c}\varphi(z,u-c)=w^c (1-\vartheta)z \,\Phi(w,z)
\end{split}
\end{align}
shifting the summation index by $c+1$ and again applying the definition of $\Phi(w,z)$.
By combining and rearranging $(a)$, $(b)$, and $(c)$, we can isolate $\Phi(w,z)$, so as to obtain
\begin{equation}\label{eq:capPhi}
 \Phi(w,z) =\frac{(1-\vartheta)z\left(\frac{w-w^{c+1}}{1-w}\right)-\vartheta z \varphi(z,1)}{1-(\vartheta z)/w -w^c z (1-\vartheta)}.
\end{equation}

What remains is to identify $\varphi(z,1)$ for each $z\in(-1,1)$. The key principle is that, because their ratio is a finite number, any root of the denominator in \eqref{eq:capPhi} must also be a root of the numerator. Indeed, supposing there is a unique $w(z)\in (-1,1)$ which yields a root of the denominator, this would imply that 
\begin{equation}
  \varphi(z,1) = \frac{1-\vartheta}{\vartheta} z \left( \frac{w(z)-w(z)^{c+1}}{1-w(z)}\right)=\frac{(1-\vartheta)zw(z)+\vartheta z-w(z)}{1-w(z)}.
\end{equation}
This means that we are done if we can show that for any $z\in(-1,1)$ the denominator in (\ref{eq:capPhi}) has a unique root $w(z)$ in $(-1,1).$ It is directly seen that this root satisfies 
\begin{align}
  F(w)&:= w^{c+1} z (1-\vartheta) + z \vartheta= w =:G(w).
\end{align}
Observe that for $z=0$ we obviously have the unique root $w=0.$

\textit{Subcase $z>0$.} Observe that $F(0)=z\vartheta > 0 =G(0)$ and $F(1)=z<1=G(1)$, so that $F(w)-G(w)$ changes sign an odd number of times for $w\in(0,1)$. But because $F(\cdot)$ is convex and $G(\cdot)$ is linear, the function $F(\cdot)-G(\cdot)$ changes sign at most twice. By combining the above, we conclude the existence of the unique root $w(z)\in(-1,1)$ (which actually lies between $0$ and $1$).

\textit{Subcase $z<0$.} The case of $c$ odd works analogously to the subcase $z>0$: 
$F(0)=z\vartheta < 0 =G(0)$ and $F(-1)=z>-1=G(-1)$. From the concavity of $F(\cdot)$, the existence of a unique root $w(z)\in(-1,1)$ follows (which lies between $-1$ and $0$). In case $c$ is even, we still have $F(0)=z\vartheta < 0 =G(0)$, but now it should be noted that $F(-1)=z(2\vartheta-1)$, which is, for $z\in(-1,0)$ and $\vartheta\in(0,1)$, larger than $-1=G(-1)$ (pick $\vartheta =1$ and $z=-1$). The existence of a unique root $w(z)\in(-1,1)$ now follows from the fact that $F(w)$ is decreasing and $G(w)$ is increasing for $w\in(-1,0)$ (where this root lies between $-1$ and $0$).
\end{proof}

\section{Appendix: Single agent case \texorpdfstring{$r,c\in\mathbb{N}$}{r,c in N}}\label{sec:pncn}
In this case no exact formula can be obtained for the probability of quitting or the expected time to quitting conditioned thereon; recall that in the other two cases we exploited the fact that the walk, either in the upward or the downward direction, had step size 1. We briefly sketch a numerical scheme which allows accurate approximation of the probability of quitting. The relationship for all $u>c$ is:
\begin{equation}\label{eq:system}
  \pi(u) = \vartheta\pi(u+r) + (1-\vartheta)\pi(u-c).
\end{equation}
We supplement this with an approximation for large enough $u>u_0$ by setting $\hat{\pi}(u)=\eta \gamma^u$ for $\eta\geq 0$ and $\gamma\in (0,1)$. In combination with (\ref{eq:system}) we know that $\gamma$ satisfies $1=(1-\vartheta) \gamma^{-c}+\vartheta\,\gamma^{r}$. Using this approximation for $\pi(u)$ from $u\geq u_0$ one has:
\begin{equation}\label{eq:linsys}
  \hat{\pi}(u) = \begin{cases}
  \eta \gamma^{u} & \text{if } u\geq u_0\\
  (1-\vartheta)\hat{\pi}(u-c) + \vartheta \hat{\pi}(u+r) & \text{if }c<u<u_0\\
  (1-\vartheta) + \vartheta \hat{\pi}(u+r) &\text{if } 0 <u\leq c,
  \end{cases}
\end{equation}
which can be used to construct a system of equations which can be solved numerically. We propose approximating the probability of absorption (and subsequently a termination of trust) by the solution to the system (\ref{eq:linsys}) at the appropriate $u = r\alpha-c\beta+1$. The value of $u_0$ should be large enough value to get a good approximation, and small enough to keep its computation manageable.

We sketch a similar approximation scheme for the expected time in the system. The crux of the scheme is to approximate the expected time to quitting at a high enough $u$ by the number of rounds it would take to move directly toward absorption without taking any steps away: $\lceil u/c\rceil$. In terms of $\varphi (z,u)$ one arrives at the following system of equations:
\begin{equation}\label{eq:linsysTime}
\hat{\varphi}(z,u)=\begin{cases}
(1-\vartheta)^{\lceil u/c\rceil}z^{\lceil u/c\rceil} & \text{if }u\geq u_0\\
(1-\vartheta)z\, \varphi(z,u-c) + \vartheta z\, \varphi(z,u+r) & \text{if } c< u < u_0\\
(1-\vartheta)z + \vartheta z \,\varphi(z,u+r) & \text{if } 0<u\leq c.
\end{cases}
\end{equation}
This defines a system of equations similar to (\ref{eq:linsys}) which can be solved numerically. This provides an approximation to $\varphi(z,u_\text{crit})$ which in turn can be used to get an approximation of $\mathbb{E}[\tau(u_\text{crit})\,|\,\tau(u_\text{crit})<\infty]$ by taking the derivative at $u_\text{crit}:$ $\varphi'(1,u_\text{crit})/\varphi(1,u_\text{crit})$. Notice the similarity to (\ref{eq:Etau}).

\section{\texorpdfstring{Appendix: An illustration of how the OA model can beat the OR model}{How the OA can beat the OR model}}\label{AX:illustration}
In this appendix we elaborate on the peculiar outcome that in the parameter setting with $c=3$, $r=2$, and $u^*=2$ the OA model yields a lower probability of quitting than the OR model at the trustworthiness settings $\vartheta=0.65,0.66$. 

We consider a pair of agents in precisely the setting in with the exception ($c=3$, $r=2$, $u^*=2$ and $\vartheta = 0.65$). Furthermore, we presume that of the first four times trust is placed in the institution (by either agent) trust is honoured only once. 

In the OR model, this means that maximally two rounds of interaction can take place. Either the first round includes abuses of trust to both agents, in which case they quit in round 1, or trust is honoured once in round 1, but abused twice in round 2 and so the agents do not place trust in round 3.

In the OA model, the agents involved have two rounds of interaction regardless of their first reward. The act of placing trust at the start of round 2 yields no information to the neighbour. At the start of round 3 however, the agent (labelled 1) who observed two abuses does not place trust. This signals to the neighbouring agent (labelled 2) that two abuses of trust have certainly occurred. At the point of placing trust in round 3 (before observing the outcome thereof) agent 2 has access to information which brings their estimate below the critical value. However, the agent has already placed their trust and so will see the outcome of that interaction. This outcome is of crucial importance because if the trust is honoured (which is more likely as $\vartheta=0.65>0.5$), agent 2 will stay in the relationship while if the trust is abused they quit. 

\section{Appendix: Tables of simulated results} \label{AX:tables}
In the tables below we present the probability of quitting and the expected time to quitting under the three considered mechanisms. 
\begin{table}[htb]
\centering
\caption{Probability of quitting and the expected time to quitting in the single agent model (Sol), the observable actions model (OA) and the observable rewards model (OR). Parameters $\alpha$ and $\beta$ set such that $u^*=1$}
\scalebox{0.79}{\subfloat[{\large Probability of quitting}]{
\begin{tabular}{l|rrrrrrrrrrrrr}
\multicolumn{6}{l}{$\bm{c=r=1}$} & \multicolumn{8}{r}{$\bm{\alpha = \beta =2}$}  \\
\hline
{$\vartheta$} & 0.18 &  0.24 &  0.30 &  0.36 &  0.42 &   0.45 &   0.55 &   0.60 &   0.66 &   0.72 &   0.78 &   0.84 &   0.90 \\
\hline
Sol &   1.0  &   1.0 &   1.0 &   1.0 &   1.0 &  1.000 &  0.808 &  0.669 &  0.518 &  0.390 &  0.288 &  0.198 &  0.119 \\
OA  &   1.0  &   1.0 &   1.0 &   1.0 &   1.0 &  0.999 &  0.797 &  0.649 &  0.493 &  0.370 &  0.272 &  0.186 &  0.113 \\
OR  &   1.0  &   1.0 &   1.0 &   1.0 &   1.0 &  1.000 &  0.648 &  0.429 &  0.248 &  0.138 &  0.072 &  0.030 &  0.010 \\
\hline
 \multicolumn{10}{l}{$\bm{c<r}$:}\\
\hline
 \multicolumn{6}{l}{$\bm{c=2, r=3}$} & \multicolumn{8}{r}{$\bm{\alpha =3,\beta =4}$}  \\
\hline
{$\vartheta$ }&  0.18  &  0.24 &  0.3 &  0.35 &   0.45 &   0.48 &   0.54 &    0.6 &   0.66 &   0.72 &   0.78 &   0.84 &    0.9 \\
\hline
Sol &   1.0  &   1.0 &  1.0 &   1.0 &  0.832 &  0.755 &  0.611 &  0.495 &  0.394 &  0.311 &  0.241 &  0.170 &  0.108 \\
OA  &   1.0  &   1.0 &  1.0 &   1.0 &  0.825 &  0.734 &  0.592 &  0.483 &  0.382 &  0.296 &  0.230 &  0.165 &  0.102 \\
OR  &   1.0  &   1.0 &  1.0 &   1.0 &  0.745 &  0.621 &  0.424 &  0.292 &  0.190 &  0.120 &  0.066 &  0.030 &  0.010 \\
\hline
 \multicolumn{6}{l}{$\bm{c=1, r=2}$} & \multicolumn{8}{r}{$\bm{\alpha =2,\beta =4}$}  \\
 \hline
{$\vartheta$} &   0.18 &  0.24 &  0.28 &   0.38 &   0.42 &   0.48 &   0.54 &    0.6 &   0.66 &   0.72 &   0.78 &   0.84 &    0.9 \\
\hline
Sol &   1.0  &   1.0 &   1.0 &  0.869 &  0.777 &  0.658 &  0.551 &  0.459 &  0.377 &  0.302 &  0.234 &  0.169 &  0.108 \\
OA  &   1.0   &   1.0 &   1.0 &  0.866 &  0.766 &  0.641 &  0.538 &  0.447 &  0.366 &  0.289 &  0.225 &  0.164 &  0.102 \\
OR  &   1.0   &   1.0 &   1.0 &  0.787 &  0.644 &  0.478 &  0.336 &  0.243 &  0.162 &  0.105 &  0.060 &  0.028 &  0.010 \\
\hline 
 \multicolumn{10}{l}{$\bm{c>r}$:}\\
\hline
\multicolumn{6}{l}{$\bm{c=3, r=2}$} & \multicolumn{8}{r}{$\bm{\alpha =4,\beta =2}$}  \\
\hline
{$\vartheta$} &   0.18 &  0.24 &  0.3 &  0.36 &  0.42 &  0.48 &  0.54 &   0.65 &   0.66 &   0.72 &   0.78 &   0.84 &    0.9 \\
\hline
Sol &   1.0  &   1.0 &  1.0 &   1.0 &   1.0 &   1.0 &   1.0 &  0.743 &  0.703 &  0.487 &  0.330 &  0.215 &  0.124 \\
OA  &   1.0  &   1.0 &  1.0 &   1.0 &   1.0 &   1.0 &   1.0 &  0.721 &  0.682 &  0.475 &  0.318 &  0.205 &  0.120 \\
OR  &   1.0  &   1.0 &  1.0 &   1.0 &   1.0 &   1.0 &   1.0 &  0.657 &  0.600 &  0.333 &  0.168 &  0.066 &  0.024 \\
\hline
\multicolumn{6}{l}{$\bm{c=2, r=1}$} & \multicolumn{8}{r}{$\bm{\alpha =5,\beta =2}$}  \\
\hline
{$\vartheta$} &   0.18 &  0.24 &  0.3 &  0.36 &  0.42 &  0.48 &  0.54 &  0.6 &   0.62 &   0.72 &   0.78 &  0.84 &    0.9 \\
\hline
Sol &   1.0  &   1.0 &  1.0 &   1.0 &   1.0 &   1.0 &   1.0 &  1.0 &  0.999 &  0.682 &  0.438 &  0.27 &  0.142 \\
OA  &   1.0   &   1.0 &  1.0 &   1.0 &   1.0 &   1.0 &   1.0 &  1.0 &  1.000 &  0.667 &  0.433 &  0.26 &  0.140 \\
OR  &   1.0   &   1.0 &  1.0 &   1.0 &   1.0 &   1.0 &   1.0 &  1.0 &  1.000 &  0.622 &  0.352 &  0.17 &  0.068 \\
\hline
\end{tabular}\label{tab:Qu1}}}\hfill
\scalebox{0.79}{\subfloat[{\large Expected time to quit}]{
\begin{tabular}{l|rrrrrrrrrrrrr}
\multicolumn{5}{l}{$\bm{c=r=1}$} & \multicolumn{9}{r}{$\bm{\alpha =2,\beta =2}$}  \\
\hline
{$\vartheta$} &   0.18 &   0.24 &   0.30 &   0.36 &   0.42 &    0.45 &   0.55 &   0.60 &   0.66 &   0.72 &   0.78 &   0.84 &   0.90 \\
\hline
Sol &  1.550 &  1.928 &  2.546 &  3.640 &  6.702 &  10.154 &  9.236 &  4.841 &  3.159 &  2.184 &  1.713 &  1.510 &  1.252 \\
OA  &  1.352 &  1.597 &  2.037 &  2.740 &  4.718 &   7.710 &  7.116 &  3.891 &  2.480 &  1.883 &  1.478 &  1.237 &  1.155 \\
OR  &  1.562 &  1.905 &  2.482 &  3.701 &  6.852 &  11.253 &  9.882 &  5.529 &  3.193 &  2.177 &  1.710 &  1.426 &  1.095 \\
\hline
 \multicolumn{9}{l}{$\bm{c<r}:$}\\
\hline
\multicolumn{5}{l}{$\bm{c=2, r=3}$} & \multicolumn{9}{r}{$\bm{\alpha =3,\beta =4}$}  \\
\hline
{$\vartheta$} &   0.18 &   0.24 &    0.3 &   0.35 &   0.45 &   0.48 &   0.54 &    0.6 &   0.66 &   0.72 &   0.78 &   0.84 &    0.9 \\
\hline
Sol &  1.897 &  2.711 &  4.408 &  9.158 &  8.613 &  5.552 &  3.033 &  2.136 &  1.619 &  1.388 &  1.262 &  1.089 &  1.007 \\
OA  &  1.690 &  2.231 &  3.584 &  7.342 &  7.082 &  4.439 &  2.702 &  2.057 &  1.532 &  1.308 &  1.184 &  1.121 &  1.039 \\
OR  &  1.668 &  2.197 &  3.329 &  7.241 &  7.834 &  4.925 &  2.882 &  2.096 &  1.855 &  1.583 &  1.397 &  1.339 &  1.095 \\
\hline
\multicolumn{5}{l}{$\bm{c=1, r=2}$} & \multicolumn{9}{r}{$\bm{\alpha =2,\beta =4}$}  \\
\hline
{$\vartheta$} &   0.18 &   0.24 &   0.28 &   0.38 &   0.42 &   0.48 &   0.54 &    0.6 &   0.66 &   0.72 &   0.78 &   0.84 &    0.9 \\
\hline
Sol &  2.130 &  3.558 &  6.052 &  6.755 &  4.045 &  2.690 &  1.960 &  1.541 &  1.352 &  1.243 &  1.112 &  1.053 &  1.007 \\
OA  &  1.898 &  2.904 &  5.278 &  6.204 &  3.804 &  2.277 &  1.770 &  1.463 &  1.282 &  1.192 &  1.102 &  1.084 &  1.039 \\
OR  &  1.874 &  2.975 &  4.855 &  6.630 &  3.898 &  2.506 &  1.932 &  1.724 &  1.517 &  1.352 &  1.281 &  1.218 &  1.095 \\
\hline
\multicolumn{9}{l}{$\bm{c>r:}$}\\
\hline
\multicolumn{5}{l}{$\bm{c=3, r=2}$} & \multicolumn{9}{r}{$\bm{\alpha =4,\beta =2}$}  \\
\hline
{$\vartheta$} &   0.18 &   0.24 &    0.3 &   0.36 &   0.42 &   0.48 &    0.54 &    0.65 &    0.66 &   0.72 &   0.78 &   0.84 &    0.9 \\
\hline
Sol &  1.514 &  1.824 &  2.250 &  2.831 &  3.856 &  5.650 &  11.644 &  12.538 &  10.590 &  5.090 &  2.762 &  1.988 &  1.497 \\
OA  &  1.328 &  1.531 &  1.833 &  2.242 &  2.941 &  4.552 &  10.004 &   9.398 &   7.988 &  4.010 &  2.388 &  1.750 &  1.418 \\
OR  &  1.470 &  1.706 &  2.014 &  2.476 &  3.161 &  4.583 &   9.070 &  10.020 &   8.075 &  4.647 &  3.172 &  2.500 &  2.250 \\
\hline
\multicolumn{5}{l}{$\bm{c=2, r=1}$} & \multicolumn{9}{r}{$\bm{\alpha =5,\beta =2}$}  \\
\hline
{$\vartheta$} &   0.18 &   0.24 &    0.3 &   0.36 &   0.42 &   0.48 &   0.54 &     0.6 &    0.62 &    0.72 &   0.78 &   0.84 &    0.9 \\
\hline
Sol &  1.444 &  1.692 &  2.004 &  2.414 &  3.034 &  3.958 &  5.837 &  10.977 &  15.353 &  12.359 &  5.390 &  3.174 &  1.961 \\
OA  &  1.270 &  1.418 &  1.627 &  1.891 &  2.332 &  3.072 &  4.684 &   9.232 &  13.028 &   9.437 &  4.308 &  2.448 &  1.734 \\
OR  &  1.360 &  1.522 &  1.683 &  1.954 &  2.350 &  3.007 &  4.348 &   7.880 &  11.248 &   9.034 &  4.131 &  3.009 &  2.336 \\
\hline
\end{tabular}\label{tab:Tu1}}}
\end{table}

\begin{table}[ht]
\centering
\caption{Probability of quitting and the expected time to quitting in the single agent model (Sol), the observable actions model (OA) and the observable rewards model (OR). Parameters $\alpha$ and $\beta$ set such that $u^*=2$}
\scalebox{0.79}{\subfloat[{\large Probability of quitting}]{
\begin{tabular}{l|rrrrrrrrrrrrr}
\multicolumn{6}{l}{$\bm{c=r=1}$} & \multicolumn{8}{r}{$\bm{\alpha =3, \beta =2}$}  \\
\hline
{$\vartheta$} & 0.18 &  0.24 &  0.3 &  0.36 &  0.42 &   0.45 &   0.55 &    0.6 &   0.66 &   0.72 &   0.78 &   0.84 &    0.9 \\
\hline
Sol &   1.0  &   1.0 &  1.0 &   1.0 &   1.0 &  1.000 &  0.665 &  0.450 &  0.267 &  0.159 &  0.086 &  0.038 &  0.012 \\
OA  &   1.0  &   1.0 &  1.0 &   1.0 &   1.0 &  0.999 &  0.647 &  0.429 &  0.252 &  0.141 &  0.078 &  0.034 &  0.012 \\
OR  &   1.0  &   1.0 &  1.0 &   1.0 &   1.0 &  1.000 &  0.648 &  0.429 &  0.248 &  0.138 &  0.072 &  0.030 &  0.010 \\
\hline
\multicolumn{11}{l}{$\bm{c<r:}$} \\
\hline
\multicolumn{6}{l}{$\bm{c=2, r=3}$} & \multicolumn{8}{r}{$\bm{\alpha =3, \beta =3}$}  \\
\hline
{$\vartheta$} & 0.18 &  0.24 &  0.3 &  0.35 &   0.45 &   0.48 &   0.54 &    0.6 &   0.66 &   0.72 &   0.78 &   0.84 &    0.9 \\
\hline
Sol &   1.0  &   1.0 &  1.0 &   1.0 &  0.700 &  0.574 &  0.376 &  0.249 &  0.160 &  0.101 &  0.060 &  0.028 &  0.011 \\
OA  &   1.0  &   1.0 &  1.0 &   1.0 &  0.664 &  0.532 &  0.356 &  0.228 &  0.146 &  0.089 &  0.052 &  0.025 &  0.010 \\
OR  &   1.0  &   1.0 &  1.0 &   1.0 &  0.659 &  0.526 &  0.328 &  0.204 &  0.124 &  0.082 &  0.048 &  0.022 &  0.010 \\
\hline
\multicolumn{6}{l}{$\bm{c=1, r=2}$} & \multicolumn{8}{r}{$\bm{\alpha =2, \beta =3}$}  \\
\hline
{$\vartheta$} & 0.18 &  0.24 &  0.28 &   0.38 &   0.42 &   0.48 &   0.54 &    0.6 &   0.66 &   0.72 &   0.78 &   0.84 &    0.9 \\
\hline
Sol &   1.0  &   1.0 &   1.0 &  0.760 &  0.604 &  0.434 &  0.303 &  0.214 &  0.145 &  0.096 &  0.058 &  0.028 &  0.011 \\
OA  &   1.0  &   1.0 &   1.0 &  0.747 &  0.589 &  0.418 &  0.297 &  0.204 &  0.136 &  0.086 &  0.051 &  0.025 &  0.010 \\
OR  &   1.0  &   1.0 &   1.0 &  0.724 &  0.559 &  0.386 &  0.258 &  0.176 &  0.115 &  0.080 &  0.048 &  0.022 &  0.010 \\
\hline
\multicolumn{11}{l}{$\bm{c>r:}$} \\
\hline
\multicolumn{6}{l}{$\bm{c=3, r=2}$} & \multicolumn{8}{r}{$\bm{\alpha =7, \beta =3}$}  \\
\hline
{$\vartheta$} & 0.18 &  0.24 &  0.3 &  0.36 &  0.42 &  0.48 &  0.54 &   0.65 &   0.66 &   0.72 &   0.78 &   0.84 &    0.9 \\
\hline
Sol &   1.0  &   1.0 &  1.0 &   1.0 &   1.0 &   1.0 &   1.0 &  0.572 &  0.508 &  0.241 &  0.115 &  0.047 &  0.014 \\
OA  &   1.0  &   1.0 &  1.0 &   1.0 &   1.0 &   1.0 &   1.0 &  0.526 &  0.463 &  0.224 &  0.104 &  0.039 &  0.013 \\
OR  &   1.0  &   1.0 &  1.0 &   1.0 &   1.0 &   1.0 &   1.0 &  0.542 &  0.481 &  0.208 &  0.088 &  0.032 &  0.010 \\
\hline
\multicolumn{6}{l}{$\bm{c=2, r=1}$} & \multicolumn{8}{r}{$\bm{\alpha =7, \beta =2}$}  \\
\hline
{$\vartheta$} & 0.18 &  0.24 &  0.3 &  0.36 &  0.42 &  0.48 &  0.54 &  0.6 &   0.62 &   0.72 &   0.78 &   0.84 &    0.9 \\
\hline
Sol &   1.0  &   1.0 &  1.0 &   1.0 &   1.0 &   1.0 &   1.0 &  1.0 &  0.998 &  0.482 &  0.202 &  0.078 &  0.022 \\
OA  &   1.0  &   1.0 &  1.0 &   1.0 &   1.0 &   1.0 &   1.0 &  1.0 &  0.999 &  0.466 &  0.197 &  0.072 &  0.021 \\
OR  &   1.0  &   1.0 &  1.0 &   1.0 &   1.0 &   1.0 &   1.0 &  1.0 &  1.000 &  0.441 &  0.164 &  0.048 &  0.013 \\
\hline
\end{tabular}\label{tab:u2p}}}\hfill
\scalebox{0.79}{\subfloat[{\large Expected time to quit}]{
\begin{tabular}{l|rrrrrrrrrrrrr}
\multicolumn{5}{l}{$\bm{c=r=1}$} & \multicolumn{9}{r}{$\bm{\alpha =3,\beta =2}$}  \\
\hline
{$\vartheta$} &   0.18 &   0.24 &    0.3 &   0.36 &    0.42 &    0.45 &    0.55 &    0.6 &   0.66 &   0.72 &   0.78 &   0.84 &    0.9 \\
\hline
Sol &  3.092 &  3.818 &  4.970 &  7.224 &  12.957 &  20.029 &  18.126 &  9.550 &  6.213 &  4.431 &  3.455 &  3.000 &  2.130 \\
OA  &  2.615 &  3.077 &  3.863 &  5.228 &   9.552 &  15.800 &  14.165 &  7.130 &  4.683 &  3.664 &  3.172 &  2.597 &  2.362 \\
OR  &  1.562 &  1.905 &  2.482 &  3.701 &   6.852 &  11.253 &   9.882 &  5.529 &  3.193 &  2.177 &  1.710 &  1.426 &  1.095 \\
\hline
 \multicolumn{9}{l}{$\bm{c<r}:$}\\
\hline
\multicolumn{5}{l}{$\bm{c=2, r=3}$} & \multicolumn{9}{r}{$\bm{\alpha =3,\beta =3}$}  \\
\hline
{$\vartheta$} &   0.18 &   0.24 &    0.3 &    0.35 &    0.45 &    0.48 &   0.54 &    0.6 &   0.66 &   0.72 &   0.78 &   0.84 &    0.9 \\
\hline
Sol &  3.728 &  5.220 &  8.389 &  17.154 &  16.696 &  11.193 &  6.267 &  4.447 &  3.353 &  2.737 &  2.432 &  2.126 &  2.000 \\
OA  &  3.060 &  4.100 &  6.736 &  14.250 &  13.397 &   7.572 &  5.232 &  3.762 &  2.888 &  2.511 &  2.203 &  2.148 &  2.026 \\
OR  &  2.146 &  3.056 &  4.791 &  10.426 &  10.130 &   6.521 &  3.615 &  2.262 &  1.634 &  1.323 &  1.146 &  1.068 &  1.000 \\
\hline
\multicolumn{5}{l}{$\bm{c=1, r=2}$} & \multicolumn{9}{r}{$\bm{\alpha =2,\beta =3}$}  \\
\hline
{$\vartheta$} &   0.18 &   0.24 &    0.28 &    0.38 &   0.42 &   0.48 &   0.54 &    0.6 &   0.66 &   0.72 &   0.78 &   0.84 &    0.9 \\
\hline
Sol &  4.173 &  7.032 &  11.962 &  14.484 &  8.156 &  5.432 &  3.862 &  3.085 &  2.767 &  2.584 &  2.258 &  2.082 &  2.000 \\
OA  &  3.364 &  5.240 &   9.371 &  11.275 &  6.624 &  4.143 &  3.223 &  2.668 &  2.415 &  2.311 &  2.138 &  2.091 &  2.026 \\
OR  &  2.384 &  3.978 &   6.820 &   8.322 &  5.202 &  2.833 &  1.959 &  1.586 &  1.300 &  1.189 &  1.146 &  1.068 &  1.000 \\
\hline
\multicolumn{9}{l}{$\bm{c>r:}$}\\
\hline
\multicolumn{5}{l}{$\bm{c=3, r=2}$} & \multicolumn{9}{r}{$\bm{\alpha =7,\beta =3}$}  \\
\hline
{$\vartheta$} &   0.18 &   0.24 &    0.3 &   0.36 &   0.42 &    0.48 &    0.54 &    0.65 &    0.66 &   0.72 &   0.78 &   0.84 &    0.9 \\
\hline
Sol &  2.992 &  3.560 &  4.337 &  5.489 &  7.374 &  10.940 &  22.088 &  25.305 &  20.474 &  9.427 &  5.633 &  4.096 &  3.000 \\
OA  &  2.559 &  2.906 &  3.429 &  4.159 &  5.572 &   8.398 &  17.574 &  16.967 &  14.051 &  7.563 &  4.701 &  3.219 &  2.755 \\
OR  &  1.551 &  1.869 &  2.298 &  3.072 &  4.174 &   6.394 &  12.963 &  13.472 &  11.554 &  5.665 &  2.780 &  1.656 &  1.095 \\
\hline
\multicolumn{5}{l}{$\bm{c=2, r=1}$} & \multicolumn{9}{r}{$\bm{\alpha =7,\beta =2}$}  \\
\hline
{$\vartheta$} &   0.18 &   0.24 &    0.3 &   0.36 &   0.42 &   0.48 &    0.54 &     0.6 &    0.62 &    0.72 &    0.78 &   0.84 &    0.9 \\
\hline
Sol &  2.851 &  3.288 &  3.874 &  4.686 &  5.866 &  7.657 &  11.462 &  21.216 &  30.238 &  25.010 &  10.048 &  6.758 &  4.529 \\
OA  &  2.467 &  2.712 &  3.087 &  3.552 &  4.325 &  5.639 &   8.380 &  16.212 &  23.482 &  16.732 &   7.119 &  4.745 &  3.602 \\
OR  &  1.520 &  1.794 &  2.142 &  2.666 &  3.360 &  4.418 &   6.550 &  12.495 &  17.276 &  13.907 &   5.960 &  3.320 &  1.731 \\
\hline
\end{tabular}\label{tab:u2t}}}
\end{table}

\begin{table}[ht]
\centering
\caption{Probability of quitting and expected time to quit in the single agent model (Sol), the observable actions model (OA) and the observable rewards model (OR). Parameters $\alpha$ and $\beta$ set such that $u^*=3$}
\scalebox{0.79}{\subfloat[{\large Expected time to quit}]{
\begin{tabular}{l|rrrrrrrrrrrrr}
\multicolumn{6}{l}{$\bm{c=r=1}$} & \multicolumn{8}{r}{$\bm{\alpha =4, \beta =2}$}  \\
\hline
{$\vartheta$} & 0.18 &  0.24 &  0.3 &  0.36 &  0.42 &   0.45 &   0.55 &    0.6 &   0.66 &   0.72 &   0.78 &   0.84 &    0.9 \\
\hline
Sol &   1.0 &   1.0 &  1.0 &   1.0 &   1.0 &  1.000 &  0.544 &  0.304 &  0.138 &  0.063 &  0.025 &  0.006 &  0.001 \\
OA  &   1.0 &   1.0 &  1.0 &   1.0 &   1.0 &  0.999 &  0.526 &  0.293 &  0.139 &  0.068 &  0.024 &  0.007 &  0.001 \\
OR  &   1.0 &   1.0 &  1.0 &   1.0 &   1.0 &  1.000 &  0.441 &  0.194 &  0.062 &  0.019 &  0.006 &  0.000 &  0.000 \\
\hline
\multicolumn{11}{l}{$\bm{c<r:}$}\\
\hline
\multicolumn{6}{l}{$\bm{c=2, r=3}$} & \multicolumn{8}{r}{$\bm{\alpha =5, \beta =5}$}  \\
\hline
{$\vartheta$} & 0.18 &  0.24 &  0.3 &   0.35 &   0.45 &   0.48 &   0.54 &    0.6 &   0.66 &   0.72 &   0.78 &   0.84 &    0.9 \\
\hline
Sol &   1.0 &   1.0 &  1.0 &  0.999 &  0.594 &  0.436 &  0.238 &  0.128 &  0.066 &  0.033 &  0.013 &  0.004 &  0.001 \\
OA  &   1.0 &   1.0 &  1.0 &  1.000 &  0.564 &  0.419 &  0.232 &  0.124 &  0.067 &  0.035 &  0.014 &  0.004 &  0.001 \\
OR  &   1.0 &   1.0 &  1.0 &  1.000 &  0.546 &  0.386 &  0.186 &  0.084 &  0.030 &  0.012 &  0.005 &  0.000 &  0.000 \\
\hline
\multicolumn{6}{l}{$\bm{c=1, r=2}$} & \multicolumn{8}{r}{$\bm{\alpha =2, \beta =2}$}  \\
\hline
{$\vartheta$} & 0.18 &  0.24 &  0.28 &   0.38 &   0.42 &   0.48 &   0.54 &    0.6 &   0.66 &   0.72 &   0.78 &   0.84 &    0.9 \\
\hline
Sol &   1.0 &   1.0 &   1.0 &  0.662 &  0.470 &  0.288 &  0.170 &  0.100 &  0.055 &  0.028 &  0.013 &  0.004 &  0.001 \\
OA  &   1.0 &   1.0 &   1.0 &  0.622 &  0.441 &  0.264 &  0.152 &  0.092 &  0.051 &  0.028 &  0.012 &  0.004 &  0.001 \\
OR  &   1.0 &   1.0 &   1.0 &  0.610 &  0.406 &  0.216 &  0.114 &  0.051 &  0.017 &  0.006 &  0.003 &  0.000 &  0.000 \\
\hline
\multicolumn{11}{l}{$\bm{c<r:}$} \\
\hline
\multicolumn{6}{l}{$\bm{c=3, r=2}$} & \multicolumn{8}{r}{$\bm{\alpha =7, \beta =2}$}  \\
\hline
{$\vartheta$} & 0.18 &  0.24 &  0.3 &  0.36 &  0.42 &  0.48 &  0.54 &   0.65 &   0.66 &   0.72 &   0.78 &   0.84 &    0.9 \\
\hline
Sol &   1.0 &   1.0 &  1.0 &   1.0 &   1.0 &   1.0 &   1.0 &  0.439 &  0.367 &  0.129 &  0.040 &  0.011 &  0.001 \\
OA  &   1.0 &   1.0 &  1.0 &   1.0 &   1.0 &   1.0 &   1.0 &  0.396 &  0.317 &  0.116 &  0.036 &  0.008 &  0.001 \\
OR  &   1.0 &   1.0 &  1.0 &   1.0 &   1.0 &   1.0 &   1.0 &  0.404 &  0.334 &  0.100 &  0.025 &  0.004 &  0.000 \\
\hline
\multicolumn{7}{l}{$\bm{c=2, r=1}$} & \multicolumn{4}{r}{$\bm{\alpha =9, \beta =2}$}  \\
\hline
{$\vartheta$} & 0.18 &  0.24 &  0.3 &  0.36 &  0.42 &  0.48 &  0.54 &  0.6 &   0.62 &   0.72 &   0.78 &   0.84 &    0.9 \\
\hline
Sol &   1.0 &   1.0 &  1.0 &   1.0 &   1.0 &   1.0 &   1.0 &  1.0 &  0.996 &  0.344 &  0.096 &  0.022 &  0.002 \\
OA  &   1.0 &   1.0 &  1.0 &   1.0 &   1.0 &   1.0 &   1.0 &  1.0 &  0.999 &  0.320 &  0.079 &  0.015 &  0.002 \\
OR  &   1.0 &   1.0 &  1.0 &   1.0 &   1.0 &   1.0 &   1.0 &  1.0 &  0.999 &  0.315 &  0.076 &  0.011 &  0.000 \\
\hline
\end{tabular}\label{tab:u3p}}}
\scalebox{0.79}{\subfloat[{\large Expected time to quit}]{
\begin{tabular}{l|rrrrrrrrrrrrr}
\multicolumn{5}{l}{$\bm{c=r=1}$} & \multicolumn{9}{r}{$\bm{\alpha =4,\beta =2}$}  \\
\hline
{$\vartheta$} &   0.18 &   0.24 &    0.3 &   0.36 &    0.42 &    0.45 &    0.55 &     0.6 &    0.66 &   0.72 &   0.78 &   0.84 &    0.9 \\
\hline
Sol &  3.966 &  4.680 &  5.788 &  7.518 &  10.987 &  19.082 &  29.922 &  28.213 &  15.061 &  9.271 &  6.518 &  5.260 &  4.760 \\
OA  &  3.463 &  3.834 &  4.468 &  5.503 &   7.552 &  13.513 &  22.291 &  19.470 &  10.387 &  6.623 &  5.273 &  4.385 &  3.815 \\
OR  &  2.646 &  3.140 &  3.893 &  5.038 &   7.396 &  13.353 &  21.750 &  20.367 &  11.100 &  6.712 &  4.921 &  3.454 &  4.000 \\
\hline
 \multicolumn{9}{l}{$\bm{c<r}:$}\\
\hline
\multicolumn{5}{l}{$\bm{c=2, r=3}$} & \multicolumn{9}{r}{$\bm{\alpha =5,\beta =5}$}  \\
\hline
{$\vartheta$} &   0.18 &   0.24 &    0.3 &    0.35 &    0.45 &    0.48 &    0.54 &    0.6 &   0.66 &   0.72 &   0.78 &   0.84 &    0.9 \\
\hline
Sol &  4.487 &  5.678 &  7.859 &  12.435 &  24.902 &  26.330 &  16.966 &  9.608 &  6.672 &  5.149 &  4.614 &  3.623 &  3.429 \\
OA  &  3.680 &  4.426 &  5.845 &   9.346 &  20.049 &  18.274 &  11.024 &  7.528 &  5.453 &  4.598 &  3.993 &  3.536 &  3.389 \\
OR  &  2.724 &  3.393 &  4.652 &   7.328 &  15.510 &  15.062 &  10.206 &  5.879 &  4.389 &  3.881 &  3.522 &  3.100 &  4.000 \\
\hline
\multicolumn{5}{l}{$\bm{c=1, r=2}$} & \multicolumn{9}{r}{$\bm{\alpha =2,\beta =2}$}  \\
\hline
{$\vartheta$} &   0.12 &   0.18 &    0.24 &    0.28 &    0.38 &    0.42 &   0.48 &   0.54 &    0.6 &   0.66 &   0.72 &   0.78 & 0.84\\
\hline
Sol &  4.733 &  6.418 &  10.601 &  18.015 &  21.900 &  12.759 &  8.160 &  5.753 &  4.537 &  4.059 &  3.868 &  3.471 & 3.429\\
OA  &  3.911 &  5.140 &   8.152 &  14.868 &  16.413 &  10.208 &  6.396 &  4.823 &  4.194 &  3.628 &  3.460 &  3.229 & 3.000 \\
OR  &  2.912 &  3.972 &   6.327 &  10.711 &  13.418 &   8.477 &  4.984 &  3.833 &  3.157 &  2.794 &  2.546 &  2.500 & Null \\
\hline
\multicolumn{9}{l}{$\bm{c>r:}$}\\
\hline
\multicolumn{5}{l}{$\bm{c=3, r=2}$} & \multicolumn{9}{r}{$\bm{\alpha =7,\beta =2}$}  \\
\hline
{$\vartheta$} &   0.18 &   0.24 &    0.3 &   0.36 &   0.42 &    0.48 &    0.54 &    0.65 &    0.66 &    0.72 &    0.78 &   0.84 &    0.9 \\
\hline
Sol &  3.911 &  4.511 &  5.332 &  6.422 &  8.088 &  10.913 &  16.426 &  32.571 &  36.403 &  29.336 &  13.765 &  8.543 &  7.140 \\
OA  &  3.457 &  3.799 &  4.323 &  5.068 &  6.141 &   8.104 &  12.119 &  24.871 &  25.874 &  20.474 &  10.281 &  7.240 &  4.645 \\
OR  &  2.570 &  2.910 &  3.344 &  3.907 &  4.867 &   6.274 &   9.379 &  18.112 &  19.234 &  17.288 &   8.740 &  4.880 &  4.000 \\
\hline
\multicolumn{5}{l}{$\bm{c=2, r=1}$} & \multicolumn{9}{r}{$\bm{\alpha =9,\beta =2}$}  \\
\hline
{$\vartheta$} &   0.18 &   0.24 &    0.3 &   0.36 &   0.42 &   0.48 &    0.54 &     0.6 &    0.62 &    0.72 &    0.78 &    0.84 &    0.9 \\
\hline
Sol &  3.815 &  4.305 &  4.963 &  5.782 &  6.950 &  8.716 &  11.345 &  16.756 &  31.361 &  43.763 &  37.012 &  15.414 &  9.261 \\
OA  &  3.446 &  3.740 &  4.151 &  4.683 &  5.392 &  6.558 &   8.474 &  12.322 &  23.541 &  33.944 &  25.088 &  10.330 &  7.312 \\
OR  &  2.442 &  2.682 &  2.984 &  3.358 &  4.006 &  4.910 &   6.421 &   9.459 &  17.515 &  24.158 &  20.682 &   9.183 &  4.500 \\
\hline
\end{tabular}\label{tab:u3t}}}
\end{table}
\end{document}